\documentclass[trackchanges,twocolumn]{aastex701}

\usepackage{svg}
\usepackage{caption}
\usepackage{comment}
\usepackage{multirow}
\usepackage{tabularx}
\usepackage{longtable}
\usepackage{graphicx}
\usepackage{subfigure}
\usepackage{threeparttable}
\usepackage{booktabs}

\begin{document}

\title{SN 2025fhm: A central-engine powered Ic-BL supernova associated with X-ray transient EP250304a}

\author[orcid=0000-0001-8390-9962, sname='Song']{Cui-Ying Song}
\altaffiliation{These authors contributed equally}
\affiliation{Department of Physics, Tsinghua University, Beijing 100084, China}
\email{cy6201991@163.com}

\author[0000-0001-8748-0016]{Nan Jiang}
\altaffiliation{These authors contributed equally}
\affiliation{David A. Dunlap Department of Astronomy and Astrophysics, University of Toronto, 50 St. George Street, Toronto, ON M5S 3H4, Canada}
\email{nan.jiang@astro.utoronto.ca}

\author[orcid=0000-0002-7334-2357, sname='Wang']{Xiaofeng Wang}
\affiliation{Department of Physics, Tsinghua University, Beijing 100084, China}
\email[show]{wang\_xf@tsinghua.edu.cn}

\author[0000-0002-5596-5059]{Yi-Han Iris Yin}
\affiliation{Department of Physics, University of Hong Kong, Pokfulam Road, Hong Kong, China}
\affiliation{Hong Kong Institute for Astronomy and Astrophysics, University of Hong Kong, Pokfulam Road, Hong Kong, China}
\email{iris.yh.yin@connect.hku.hk}

\author[orcid=0000-0002-1094-3817,sname='Wang' ]{Lingzhi Wang}
\affiliation{College of Science, Hainan Tropical Ocean University, Sanya 572022, China}
\affiliation{Chinese Academy of Sciences South America Center for Astronomy (CASSACA), National Astronomical Observatories, CAS, Beijing 100101, China}
\email{wanglingzhi@hntou.edu.cn}

\author[orcid=0000-0002-0096-3523,sname='Li']{Wenxiong Li}
\affiliation{National Astronomical Observatories, Chinese Academy of Sciences, Beijing 100101, China}
\email[show]{liwx@bao.ac.cn}

\author[]{Shengyu Yan}
\affiliation{Department of Physics, Tsinghua University, Beijing 100084, China}
\email{yan_shengyu.astro@outlook.com}

\author[0000-0003-4200-5064]{Dae-Sik Moon}
\affiliation{David A. Dunlap Department of Astronomy and Astrophysics, University of Toronto, 50 St. George Street, Toronto, ON M5S 3H4, Canada}
\email{daesik.moon@utoronto.ca}

\author[]{Tao An}
\affiliation{Department of Astronomy, University of Science and Technology of China, 96 Jinzhai Road, Hefei, Anhui 230026, China}
\email{antao2008@ustc.edu.cn}

\author[0000-0001-7101-9831]{Aleksandar Cikota}
\affiliation{Gemini Observatory/NSF's NOIRLab, Casilla 603, La Serena, Chile}
\email{aleksandar.cikota@noirlab.edu}

\author[0000-0002-1481-4676]{Samaporn~Tinyanont}
\affiliation{National Astronomical Research Institute of Thailand, 260 Moo 4, Donkaew, Maerim, Chiang Mai, 50180, Thailand}
\email{samaporn@narit.or.th}

\author[]{Liang-Duan Liu}
\affiliation{Institute of Astrophysics, Central China Normal University, Wuhan 430079, China}
\email{liuld@ccnu.edu.cn}

\author[0000-0002-0170-0741]{Cui-Yuan Dai}
\affiliation{School of Astronomy and Space Science, Nanjing University, Nanjing 210093, China}
\affiliation{Key Laboratory of Modern Astronomy and Astrophysics (Nanjing University), Ministry of Education, Nanjing 210023, China}
\email{cydai@smail.nju.edu.cn}

\author[0000-0001-9732-2281]{Christopher D. Matzner}
\affiliation{David A. Dunlap Department of Astronomy and Astrophysics, University of Toronto, 50 St. George Street, Toronto, ON M5S 3H4, Canada}
\email{matzner@astro.utoronto.ca}

\author[]{Bin-Bin Zhang}
\affiliation{School of Astronomy and Space Science, Nanjing University, Nanjing 210093, China}
\affiliation{Key Laboratory of Modern Astronomy and Astrophysics (Nanjing University), Ministry of Education, Nanjing 210023, China}
\email{bbzhang@nju.edu.cn}

\author[0009-0000-0023-1296]{Lixin Yu}
\affiliation{Chinese Academy of Sciences South America Center for Astronomy (CASSACA), National Astronomical Observatories, CAS, Beijing 100101, China}
\email{nixil.itp@gmail.com}

\author[]{Qinyu Wu}
\affiliation{National Astronomical Observatories, Chinese Academy of Sciences, Beijing 100101, China}
\affiliation{School of Astronomy and Space Science, University of Chinese Academy of Sciences, Beijing 100049, China}
\email{qywu@bao.ac.cn}

\author[0000-0002-3505-3036]{Hong Soo Park}
\affiliation{Korea Astronomy and Space Science Institute, 776, Daedeokdae-ro, Yuseong-gu, Daejeon 34055, Republic of Korea}
\affiliation{Korea University of Science and Technology (UST), 217 Gajeong-ro, Yuseong-gu, Daejeon 34113, Republic of Korea}

\email{hspark@kasi.re.kr}

\author[0000-0001-9670-1546]{Sang Chul Kim}
\affiliation{Korea Astronomy and Space Science Institute, 776, Daedeokdae-ro, Yuseong-gu, Daejeon 34055, Republic of Korea}
\affiliation{Korea University of Science and Technology (UST), 217 Gajeong-ro, Yuseong-gu, Daejeon 34113, Republic of Korea}\email{sckim@kasi.re.kr}

\author[0000-0002-6261-1531]{Youngdae Lee}
\affiliation{Korea Astronomy and Space Science Institute, 776, Daedeokdae-ro, Yuseong-gu, Daejeon 34055, Republic of Korea}
\email{hippo206@gmail.com}

\author[]{Yu-Hao Zhang}
\affiliation{Institute of Astrophysics, Central China Normal University, Wuhan 430079, China}
\email{zhang-yh@mails.ccnu.edu.cn}

\author[]{Haowei Peng}
\affiliation{Department of Physics, Tsinghua University, Beijing 100084, China}
\email{phw23@mails.tsinghua.edu.cn}

\author[]{Franz E. Bauer}
\affiliation{Instituto de Alta Investigación, Universidad de Tarapacá, Casilla 7D, Arica, Chile}
\email{franz.e.bauer@gmail.com}

\author[0000-0003-4914-5625]{Joseph R. Farah}
\affiliation{Las Cumbres Observatory, 6740 Cortona Dr, Goleta, CA 93117, USA}
\email{josephfarah@berkeley.edu}

\author[0000-0002-1895-6639]{Moira Andrews}
\affiliation{Las Cumbres Observatory, 6740 Cortona Dr, Goleta, CA 93117, USA}
\email{mandrews@lco.global}

\author[0009-0006-7296-728X]{Kathryn Wynn}
\affiliation{Las Cumbres Observatory, 6740 Cortona Dr, Goleta, CA 93117, USA}
\email{kwynn@lco.global}

\author[orcid=0000-0003-3656-5268, sname='Ni']{Yuan Qi Ni}
\affiliation{Kavli Institute for Theoretical Physics, University of California, Santa Barbara, 552 University Road, Goleta, 93106-4030, CA, USA} 
\affiliation{Las Cumbres Observatory, 6740 Cortona Dr, Goleta, CA 93117, USA}
\email{christopher.of.ni@gmail.com}

\author[0000-0003-4253-656X]{D. Andrew Howell}
\affiliation{Las Cumbres Observatory, 6740 Cortona Dr, Goleta, CA 93117, USA}
\affiliation{Department of Physics, University of California, Santa Barbara, CA 93106-9530, USA}
\email{ahowell@lco.global}

\author[]{Curtis McCully}
\affiliation{Las Cumbres Observatory, 6740 Cortona Dr, Goleta, CA 93117, USA}
\email{cmccully@lco.global}

\author[orcid=0000-0002-4731-9698, sname='Sun']{Ning-Chen Sun}
\affiliation{School of Astronomy and Space Science, University of Chinese Academy of Sciences, Beijing 100049, China}
\affiliation{National Astronomical Observatories, Chinese Academy of Sciences, Beijing 100101, China}
\affiliation{Institute for Frontiers in Astronomy and Astrophysics, Beijing Normal University, Beijing, 102206, China }
\email{sunnc@ucas.ac.cn}

\author[0000-0002-1089-1519]{Danfeng Xiang}
\affiliation{Beijing Planetarium, Beijing Academy of Sciences and Technology, Beijing 100044, China}
\email{xiangdanfeng2015@163.com}

\author[]{Yuan Liu}
\affiliation{National Astronomical Observatories, Chinese Academy of Sciences, Beijing 100101, China}
\email{liuyuan@bao.ac.cn}

\author[]{Wenxin Wang}
\affiliation{National Astronomical Observatories, Chinese Academy of Sciences, Beijing 100101, China}
\email{wxwang@nao.cas.cn}

\author[]{Yijia Zhang}
\affiliation{Kavli Institute for Astronomy and Astrophysics, Peking University, Beijing 100871, China}
\email{zhangyij26@pku.edu.cn}

\author[]{Wei Chen}
\affiliation{School of Astronomy and Space Science, University of Chinese Academy of Sciences, Beijing 100049, China}
\email{chenwei@bao.ac.cn}

\begin{abstract}
We present X-ray, optical, and radio follow-up observations of EP250304a, an extragalactic fast X-ray transient (EFXT) discovered by the Einstein Probe. Its X-ray light curve exhibits two broad pulses with comparable peak fluxes within the first $\sim$1~ks, a feature rarely seen among low-luminosity gamma-ray bursts or EFXTs. Optical follow-up observations were carried out using the Korea Microlensing Telescope Network, the Thai Robotic Telescope, the Las Cumbres Observatory 1~m global network, the Gemini Multi-Object Spectrograph on Gemini south telescope, and the Global Supernova Network. The fast-cooling phase (within 3 days) of optical data can be well fitted by a shocked cocoon model. However, during the supernova phase (SN 2025fhm, from 3 to 88 days), the late-time light curve cannot be explained solely by radioactive $^{56}$Ni decay, as demonstrated by a grid of simulations using the one-dimensional Lagrangian radiation hydrodynamics code SNEC, which reveals a significant energy excess at late epochs. To account for this excess, a central engine like a rapidly spinning, highly magnetized neutron star is needed to provide additional energy injection. This model yields a best-fit spin period of $\sim$12.60~ms and magnetic field strength of $\sim 3.52\times10^{15} \rm G$, and it successfully explains both the late-time bolometric light curve and the early X-ray pulse structures. Our results indicate that EP250304a/SN 2025fhm is likely powered by a central magnetar rather than by radioactive decay alone, offering new insights into the energy budget and physical origin of EFXTs and their associated supernovae.
 
\end{abstract}


\section{Introduction} 
Extragalactic fast X-ray transients (EFXTs) constitute a class of X-ray phenomena whose emission persists on timescales spanning from several seconds up to thousands of seconds. 
Although more than thirty extragalactic EFXT candidates have been reported by the Chandra, X-ray Multi-mirror Mission Newton (XMM-Newton), Swift and eROSITA observatories \citep{Eappachen2024,Quirolava2023}, only a small fraction of these events have been accompanied by extensive multi-wavelength observations. With the operation of the Einstein Probe (EP, \citealt{Yuan2022,Yuan2025}), EFXTs detection shows an obvious increase in both sample and the corresponding follow-up multiband observations. 

As one of the first EP transients to be accompanied by comprehensive multi-wavelength follow-up observations, EP240315a was linked to the gamma-ray burst (GRB) 240315C, pointing to the possibility that a fraction of EFXTs are either low-luminosity GRBs or GRBs exhibiting weak prompt emission \citep{Liu2025,Levan2025}. Meanwhile, analysis of the GRB-less source EP241021a suggests that EP FXTs may still be linked to GRBs, despite the lack of gamma-ray detections for the majority of EP transients \citep{Busmann2025}.

This interpretation is further supported by \cite{OConnor2025} who found that the redshift distributions of EFXTs detected by EP and long GRBs are consistent with a common origin. Separately, \cite{Guo2025} reported a significant correlation between the formation rates of EFXTs and long GRBs. These empirical links naturally align with the well-established progenitor framework for long GRBs. Previous observations have confirmed the link between LGRBs and broad-line type Ic supernovae \citep[SNe Ic-BL, ][]{Galama1998,Bloom1999,Macfadyen1999,Heger2003,Woosley2006,Modjaz2008}, which themselves represent a significant fraction ($\sim 18.0\%$) of the Type Ibc supernova sample \citep{Ma2025}. Core collapse of massive stars may result in a stellar-mass BH surrounded by an accretion disk \citep{Woosley1993,Macfadyen1999,Popham1999,Liu2017,Song2019,Wei2019}, or a highly magnetized, rapidly rotating neutron star (magnetar, \citealt{Usov1992,Duncan1992,Zhang2001,Metzger2011,Song2023}), producing an ultra-relativistic jet to power prompt gamma-ray emission \citep{Kumar2015}. The interaction between the jet and the circumstellar material produces multi-wavelength afterglow \citep{Sari1998,Sari1999,Meszaros2006,Zhang2006}. Thus, the similarities in redshift and formation rate between EFXTs and LGRBs strongly suggest that at least some EFXTs share a massive-star collapse origin.

Notably, only a handful of EFXT events have been found to be associated with type Ic SNe without accompanied GRBs. An example is EP240414a/SN 2024gsa, which was detected by EP but not by other GRB satellites \citep{Sun2025,Srivastav2025,Bright2025}. The evolution of the X-ray prompt emission of EP240414a exhibits a very soft energy spectrum with low peak energy $E_{\rm p} < 1.3 \rm ~keV$, distinguishing it from long GRBs or low-luminosity GRBs and X-ray flashes. The optical light curve displays at least three distinct emission phases that might corresponding to the GRB afterglow, cooling emission from the shock-heated extended material surrounding the progenitor, and a nickel-powered supernova \citep{Sun2025}. Although other models, such as the jet-cocoon system \citep{Hamidani2025,Zheng2025} and whatever mechanism powers the fast blue optical transients \citep{Van2025}, have been proposed to explain its complex light curves, it is now widely accepted that these phenomena are all associated with the death processes of massive stars.

A similar pattern of complex light-curve behavior was seen in EP250108a \citep{Li_25_25kg,Rastinejad_25_25kg,Roman_Aguilar_25_25kg,Eyles-Ferris2025}, whose optical emission exhibits multiple distinct episodes reminiscent of those in EP240414a. \citet{Li_25_25kg} found that EP250108a extends GRBs and XRFs into the unprecedentedly soft and weak regime ($E_{\rm peak} \lesssim 1.8$ keV, $E_{\rm iso} \lesssim 10^{49}$  erg), with the associated supernova SN 2025kg being one of the most luminous SNe Ic-BL ever recorded and possibly powered by a magnetar. 

More recently, EP260321a/SN 2026gzf has been identified as another member of this class. Its X-ray emission is soft and thermal, a signature often interpreted as shock breakout from the stripped-envelope progenitor \citep{Yuan2026}. \citet{Chen2026} found that enhanced mass loss shortly before core collapse produced dense circumstellar material that likely gave rise to the X-ray shock breakout signal. Deep radio and X-ray observations rule out an on-axis relativistic jet, favoring a weak or choked outflow \citep{MartinCarrillo2026,OConnor2026,Rastinejad2026,Suzuki2026}. The polarimetric observations obtained by \citet{Wen2026} reveals largely spherical outer ejecta for SN 2026gzf, implying that the explosion did not significantly disrupt the progenitor envelope.

Various progenitor models have been proposed to account for EFXTs. These include X-ray flashes or high-redshift GRBs \citep{Soderberg2004,Liu2025}, newly born millisecond magnetars \citep{Dai2006,Metzger2014,Kaspi2017,Sun2017,Xue2019,Quirola2024,Wu2025,Pang2026,Li_yufei2026}, dirty fireballs or failed GRBs \citep{Huang2002,Dai2026a,Dai2026b}, refreshed shocks in GRBs \citep{Srivastav2025,Busmann2025}, off-axis GRBs \citep{Ramirezruiz2002,Zhang2004,Piran2004,Nakar2015,Zhang2018, Wichern2024}, and two-component jets \citep{Huang2004,Jiang2025}, jet-cocoon systems \citep{Zheng2025,Hamidani2025}. Stellar flares \citep{Glennie2015}, shock breakout from core-collapse supernovae \citep{Soderberg2008,Nakar2010,Yuan2026}, tidal disruption events involving white dwarfs and intermediate-mass black holes \citep{Jonker2013,Glennie2015} have also been suggested as possible channels. Despite these extensive theoretical efforts, the physical origin of EFXTs remains highly debated, and no single model has yet succeeded in explaining the full diversity of observed properties.

Here we present the case of EP250304a and its multi-wavelength observations, together with its optical counterpart SN~2025fhm, for which no coincident GRB was reported. The structure of this paper is organized as follows: In Section 2 we introduce extensive photometric and spectroscopic data. In Section 3 we analyze our observations and compare the properties of SN~2025fhm with other type Ic-BL SNe associated with EFXTs or GRBs. In Section 4 we describe our theoretical analysis and numerical simulation models, along with constraints on progenitor stars. Discussions and conclusion are presented in Section 5.

\section{Discovery and Multiwavelength Follow-Up} \label{sec:observation_data}
\begin{figure}
    \centering
    \includegraphics[width=0.45\textwidth, trim = 0cm 9cm 0cm 0cm]{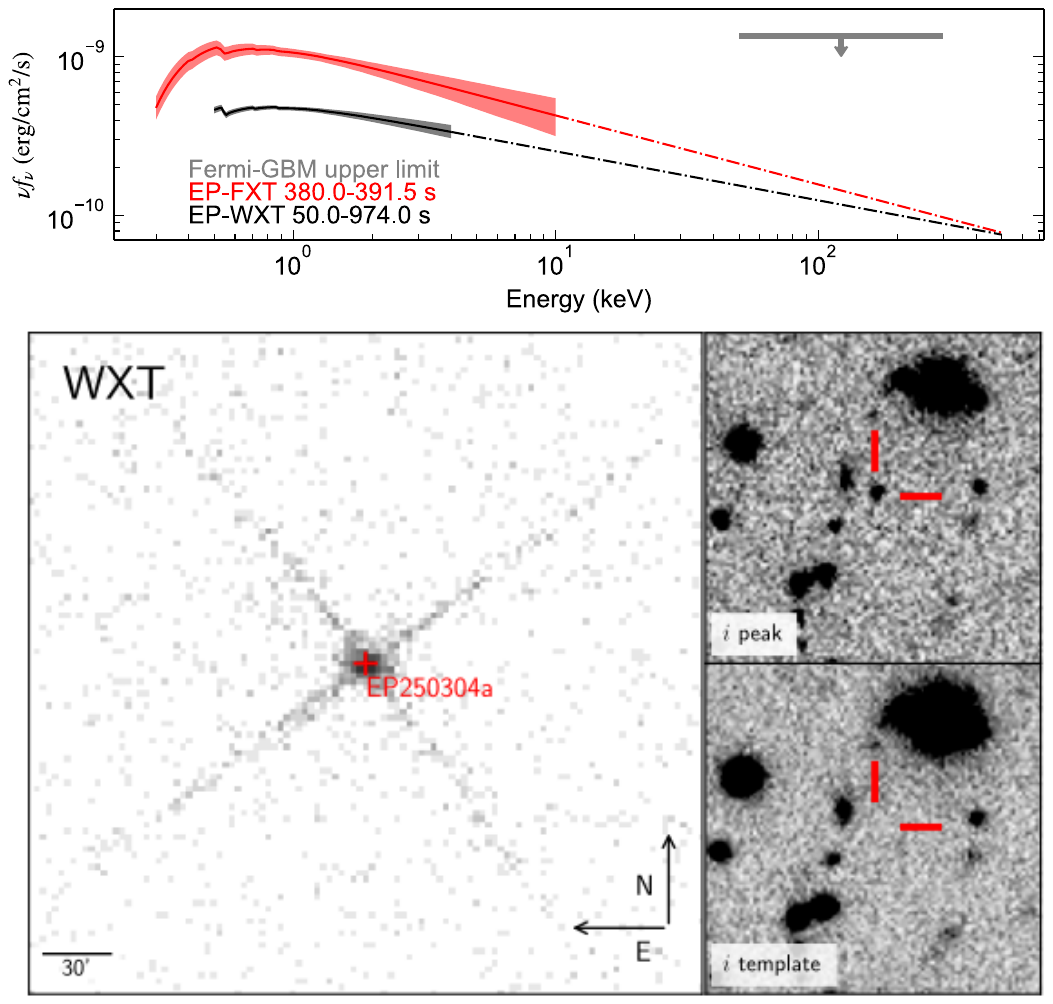} 
    \caption{\textit{Top:} The SED of EP-FXT during the peak flux interval and EP-WXT during the overlapping interval between $T_{90}$ and the GBM coverage. The solid curve denotes the SED obtained from the spectral fittings using the absorbed PL model. The dashed line represents the extrapolation of the best-fit model. The gray upper limit indicates the flux limit in the energy range of 50--300 keV derived from the Fermi-GBM data in 50.0--974.0 s. The error regions represent the 1-$\sigma$ confidence level. \textit{Bottom Left:} A $5^\circ \times 5^\circ$ zoom in view to the total photon count by WXT between Mar. 04 2025, 1 hours 34 minutes 43.885 seconds and 3 hours 30 minutes 10.817 seconds (UTC) centred at the location of the EP250304a (red cross). East is left and North is up. \textit{Bottom Right:} $i$ band 40"$\times$40" zoom in views on the field containing the location of the EP250304a (red cross) near peak (top) and template images obtained one year post explosion (bottom row). East is left and North is up.}
    \label{fig:finder}
\end{figure}

EP250304a was detected by the Wide-field X-ray Telescope (WXT) on board the Einstein Probe (EP) at 01:29:48.0 ($T_0$) on Mar.04 2025, with a position at R.A. = 13h 53m 34.58s, Dec = $-42 \rm d ~ 48\arcmin$ ~16.7$\arcsec$ (Figure \ref{fig:finder} lower panel) , and a redshift of z = 0.2 \citep{GNC_spec_redshift}. The source was followed up in the X-ray, optical, and radio bands with multiple facilities. 

\subsection{X-rays}
EP250304a triggered the on-board processing unit of the EP-WXT at $\sim T_0 + 162$ s, and the automatic follow-up observation by the EP Follow-up X-ray Telescope (FXT) began $\sim T_0 + 134$ s after the WXT trigger \citep{GCN39580, GCN39591}, following the spacecraft slew, with the two FXT modules operating in full-frame and partial-window modes, respectively. An uncatalogued fading X-ray counterpart was detected at R.A. = 208.3941 deg, Dec. = -42.8046 deg (J2000), with a positional uncertainty of 10 arcsec in radius (90\% confidence level). 

We reanalyzed the EP data published by \citet{2026arXiv260606213C}, following the standard WXT data reduction procedures (Y. Liu et al., in preparation). The resulting WXT light curve displays two broad pulses with a measured $T_{90}\sim926$ s (see also Fig. 2 of \citealt{2026arXiv260606213C}). 
The FXT data were processed using the FXT Data Analysis Software (\texttt{FXTDAS} v1.30) and the latest calibration database (CALDB v1.30). During the automatic follow-up, the FXT-B data obtained in full-frame mode were severely affected by pile-up effect and were therefore excluded from further analysis. Thus, only the FXT-A data were used in our analysis. We evaluated the pile-up effect on FXT-A following the FXTDAS user guide and removed the 30-arcsec circle region centered on the source from $T_0+296$ to $T_0+1445$ s. Within this period, the source photons were extracted from an annular region centered on the source position, with inner and outer radii of 30 and 80 arcsecs, respectively, while the background photons were extracted from a circular region with a radius of 120 arcsecs centered on a nearby clear region. For late-time data without suffering influence of pileup, circular extraction regions with radii of 60 and 120 arcsecs centered on the source and a nearby clear region, respectively, were used for the source and background, respectively.

\begin{figure*}[htb]
\centering
\includegraphics[scale=0.8]{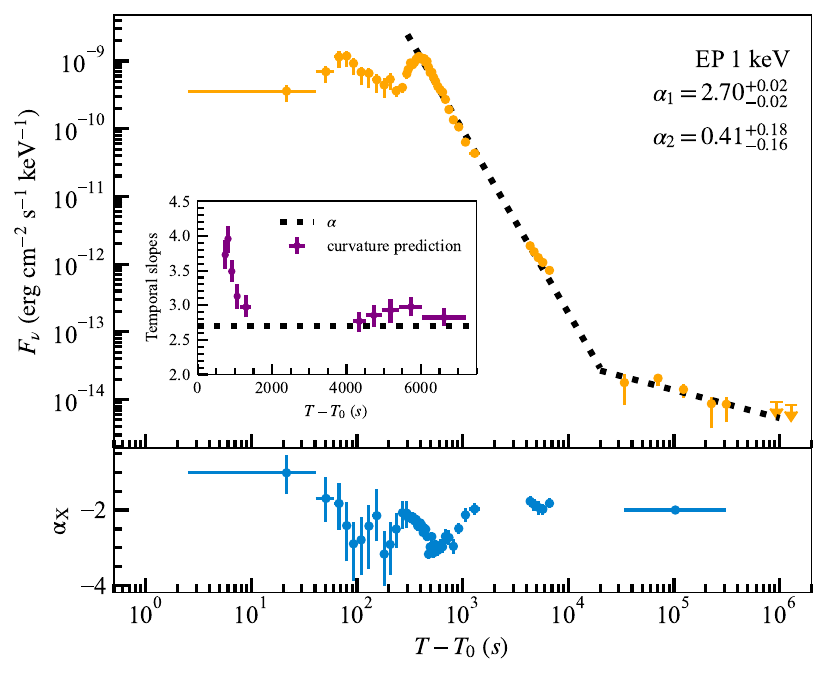}
\caption{Specific flux densities at 1 keV and the best-fit spectral indices of EP250304a observed by EP. The black dotted lines denote the best-fit power-law decays of the X-ray afterglow phases. The inset compares the observed temporal slopes during the steep decay phase with those predicted by the curvature effect. The horizontal black dotted line represents the best-fit temporal slope of the steep decay, while the data points depict the temporal slopes inferred from the curvature-effect relation. All error bars on the data points represent the 1-$\sigma$ confidence level.}
\label{fig:x_rays}
\end{figure*}

The X-ray spectra were fitted using the Bayesian-inference-based spectral fitting tool \textit{bayspec}\footnote{\url{https://github.com/jyangch/bayspec}} \citep{2022Natur.612..232Y, 2025ApJ...989L..39Y}, adopting an absorbed power-law (PL) model, \textit{tbabs*ztbabs*pl}. Here, \textit{tbabs} and \textit{ztbabs} correspond to the Tuebingen-Boulder interstellar medium absorption model \citep{Wilms2000ApJ}, accounting for the Galactic and host-frame absorption, respectively. We fixed the Galactic hydrogen absorption column density as $N_{\rm H, gal} = 5.18 \times 10^{20}$~cm$^{-2}$, while allowing the intrinsic column density $N_{\rm H, z}$ and PL parameters to vary when fitting time-integrated WXT and FXT spectra. However, the best-fit results show that both WXT and FXT spectra were unable to constrain $N_{\rm H, z}$ above $1 \times 10^{19}$~cm$^{-2}$, indicating that the host-frame absorption is negligible. We therefore adopted a simplified model, \textit{tbabs*pl}, including only the Galactic absorption component for all the spectral fits. Based on the best-fitting results, we derived the flux densities at 1 keV, which are listed in Table \ref{tab:x_rays} and illustrated in Figure \ref{fig:x_rays}. For time intervals covered simultaneously by both WXT and FXT, only the FXT measurements were retained owing to their better counting statistics. Given the limited photon statistics in the observation epochs later than $10^4$ s, we fixed the spectral index at -2 for the spectral fittings and flux estimation.

\begin{figure*}[htb]
\centering
\includegraphics[scale=0.2]{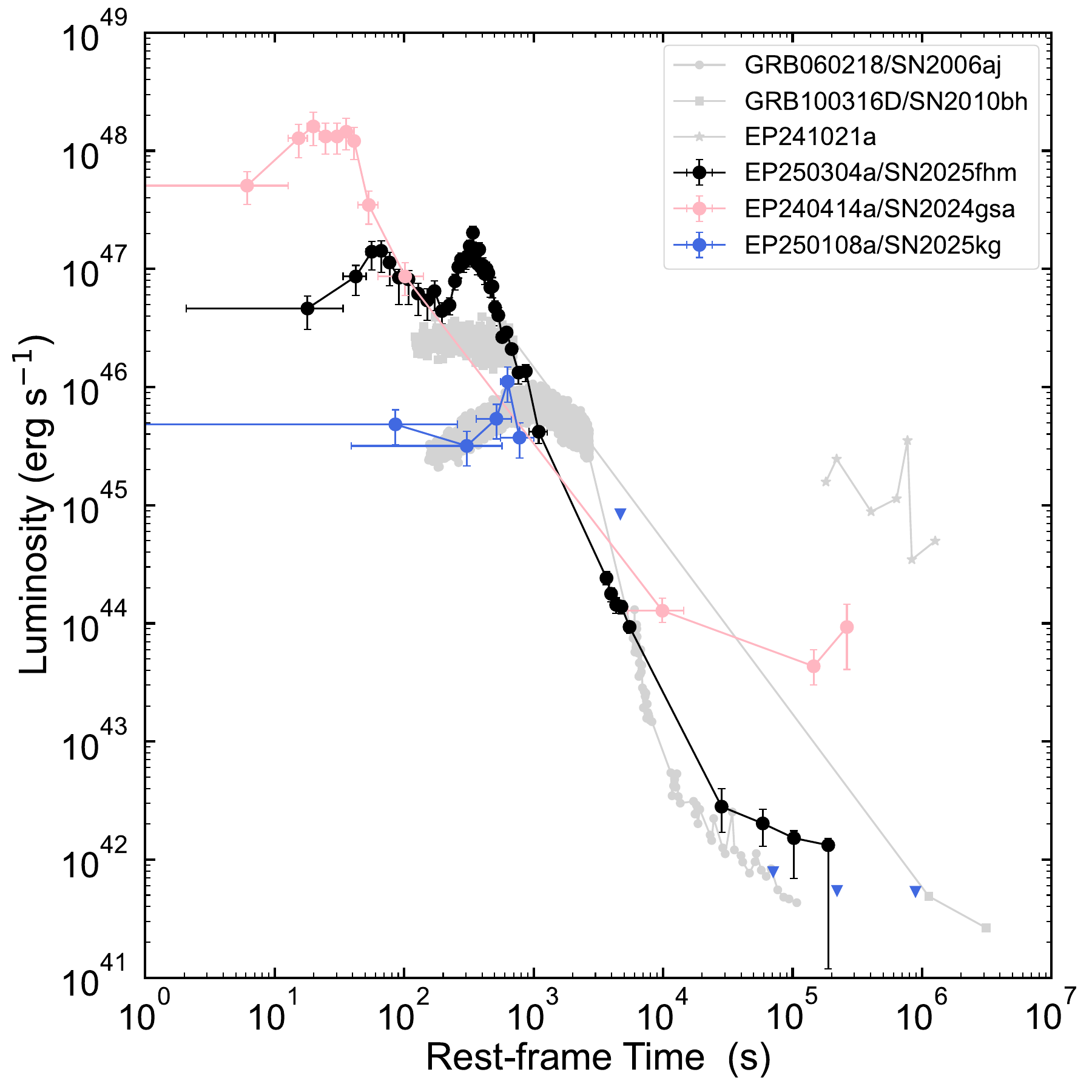}
\includegraphics[scale=0.2]{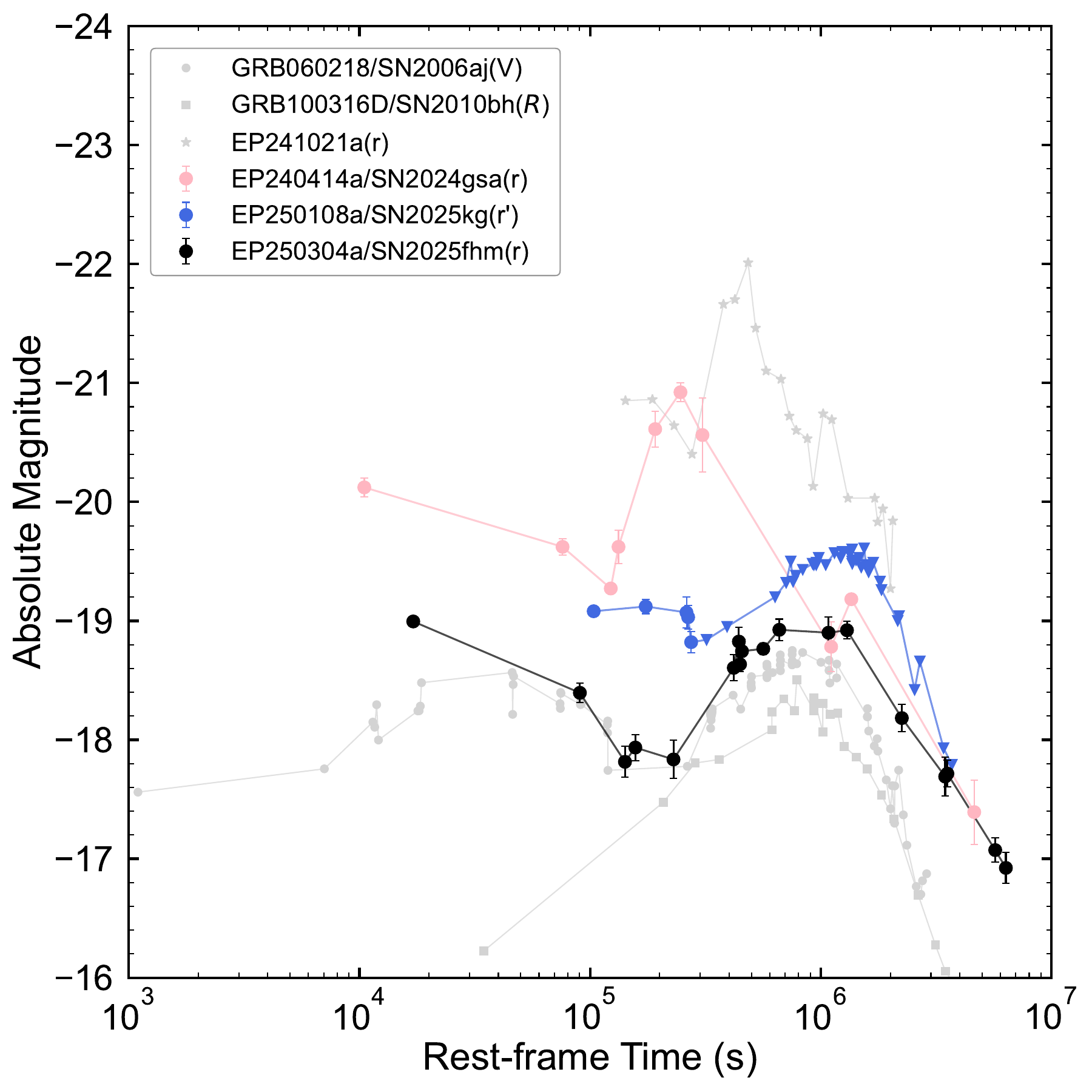}
\caption{Comparison of the X-ray (left) and optical R/r/V-band (right) light curves of EP250304a with those of other EP-detected EFXT sources and low-luminosity gamma-ray bursts.}.
\label{fig:LC_Xray_compare}
\end{figure*}

Figure \ref{fig:LC_Xray_compare} presents the temporal evolution of the X-ray luminosity of EP250304a, along with a comparison to other fast X-ray transients discovered by EP and low-luminosity gamma-ray bursts (LL-GRBs). The X-ray luminosity of EP250304a falls between those of EP240414a and EP250108a, while being notably brighter than GRB 060218 and GRB 100316D. Furthermore, its light curve exhibits a prominent double-peaked profile, a feature that was rarely seen among this class of sources.

\subsection{Gamma rays}

The prompt emission phase of EP250304a was partially covered by the Fermi Gamma-ray Burst Monitor (GBM), beginning at $\sim T_0+50$ s. However, no onboard trigger or significant ground-identified counterpart was reported in association with this event \citep{2025GCN.39600....1R}. We thereby estimated a 3$\sigma$ flux upper limit over the interval from $T_0+380$ to $T_0+391.5$ s, which encompasses the peak flux of the FXT emission, as well as the interval from $T_0+50$ to $T_0+974$ s, corresponding to the overlap between the X-ray $T_{90}$ duration and the GBM coverage. In this analysis, we selected the GBM detector n6 with the smallest viewing angle of 12 degrees to the source during these time intervals. A PL spectrum with photon index of 2 was adopted. The model normalization was determined through an iterative procedure, increasing its value until the predicted signal reached a detection significance of 3$\sigma$ above the measured background level. This procedure yields a corresponding 3$\sigma$ upper limit of $2.4\times10^{-9}~{\rm erg~cm^{-2}~s^{-1}}$ in the 50--300 keV energy band for the time interval of 50--974 s, which is above the extrapolated gamma-ray fluxes inferred from the best-fit X-ray PL spectra (Figure \ref{fig:finder}).

\subsection{Optical photometric observations}
We performed optical follow-up observations utilizing several facilities, including the 1.6~m Korea Microlensing Telescope Network (KMTNet), the 70~cm telescope of Thai-Robotic Telescope network (TRT) located at Spring Brook Remote Observatory (SBO), and Cerro Tololo Inter-American Observatory (CTO) in Chile, and the Las Cumbres Observatory 1~m global network (LC0,\cite{Brown2013}), and Gemini Multi-Object Spectrograph (GMOS) on Gemini south telescope. The above photometry spans the period from 2-3 days to about 3 months after the discovery (see Figure \ref{fig:multiband_magnetar}) and the final flux-calibrated photometric results are provided in Table 1.

\begin{deluxetable}{cccc}
\tabletypesize{\footnotesize}
\tablecolumns{4} 
%\tablewidth{0.99\textwidth}
 \tablecaption{EP250304a Apparent Magnitudes}\label{tab:photometry}
 \tablehead{
 \colhead{MJD (days)} & \colhead{Magnitude $\pm$ Error} & 
 \colhead{Band} &
 \colhead{Instrument}
 } 

\startdata 
60738.11 & 20.80 $\pm$ 0.10 & $B$ & TRT-CTO \\
60738.12 & 19.29 $\pm$ 0.09 & $U$ & Swift \\
60738.13 & 20.61 $\pm$ 0.07 & $V$ & TRT-CTO \\
60738.15 & 19.07 $\pm$ 0.06 & $U$ & Swift \\
60738.16 & 20.63 $\pm$ 0.08 & $R$ & TRT-CTO \\
60738.17 & 20.44 $\pm$ 0.17 & $I$ & TRT-CTO \\
... & ... & ... & ....
\enddata
 \tablecomments{Sample magnitudes of the optical counterpart for EP250304a without extinction nor $K$-correction.
 This table is published in its entirety in the electronic edition. A
portion is shown here for guidance regarding its formatting.}
\end{deluxetable} 

\subsubsection{KMTNet}
Under the KMTNet supernova program \citep[KSP,][]{Moon_16_ksp,2021Moon}, we conducted high-cadence, $BVi$-band follow-up observations of SN 2025fhm/EP250304a. The KMTNet consists of three identical 1.6~m telescopes located in Australia, South Africa, and Chile, which provides 24-hour continuous monitoring of the sky with a 0\farcs4 per pixel sampling \citep{Kim_16_ksp}. During the period from March 6th 2025 to April 7th 2025, we collected a total of 187 60-s and 233 120-s images reaching a 3$\sigma$-detection depth of 21--22 mag. To allow for deeper detections, we use SWARP \citep{swarp} to stack single exposures with seeing $<$ 2" taken within a maximum interval of 3.6 hours, requiring each stacked frame to be composed of at least three exposures. We then conduct aperture photometry and calibration on these stacked images with a reduced Kron radius \citep{kron} determined from fitting a Moffat function \citep{Moffat} to about 30 nearby isolated AAVSO standard stars in each band\footnote{The AAVSO Photometric All-Sky Survey: Data Release 9, https://www.aavso.org/apass} using the custom Python-based software SuperNova Analysis Package (SNAP)\footnote{\href{https://github.com/niyuanqi/SNAP}{https://github.com/niyuanqi/SNAP}}. We transform the KSP instrumental magnitudes to those of Johnson $BV$ and Sloan $i$ bands following \citet{Park_17_ksp}.

The optical counterpart of EP250304a was first detected by KSP in $B$ band at about 4 days after the EP trigger, with an apparent magnitude of 22.45$\pm$0.16 mag at the position of RA=$\rm 13^h53^m34.71^s$ and Decl.= $-42\degr48\arcmin16\farcs26$ (J2000).

In addition, we obtained 82 and 86 exposures of 60 s each in the $V$ and $i$ bands, respectively, between January 2 and March 11, 2026, corresponding to 9--12 months after the discovery of EP250304a. We combined the high-quality individual exposures using SWARP to produce deeper stacked images. No source is detected at the position of EP250304a, with $3\sigma$ upper limits of 23.02 and 23.07 mag in the $V$ and $i$ bands, respectively.

\subsubsection{TRT}
We performed $BVRI$ bands observations with the 70 cm telescope of TRT-SBO and TRT-CTO. The data were reduced by the observatory's automated pipeline. Point-spread-function (PSF) photometry was performed using \textsc{AutoPhOT}\footnote{\url{https://github.com/Astro-Sean/autophot/}} \citep{Brennan2022}. The $B$- and $V$-band photometry was calibrated against stars from the AAVSO Photometric All-Sky Survey (APASS)  catalog \citep{Henden2016}, whose Johnson $B$ and $V$ magnitudes are defined in the Vega system. For the $R$ and $I$ bands, the reference-star magnitudes were derived from the Gaia EDR3 $G$, $G_{\rm BP}$, and $G_{\rm RP}$ photometry \citep{Riello2021} using the polynomial transformations to the Johnson--Cousins $R_C$ and $I_C$ systems provided in the Gaia EDR3 documentation \citep{GaiaEDR3PhotRelations}. The resulting $BV R_C I_C$ photometry of EP250304a is reported in the Vega magnitude system.

\subsubsection{LCO}
We obtained $BVgri$-band images between March 5 and April 29, 2025 as part of the Global Supernova Project, which utilizes a network of 1.0~m LCO telescopes located at Siding Spring Observatory (Australia), the South African Astronomical Observatory (South Africa), and Cerro Tololo Inter-American Observatory (Chile). The photometric reduction and calibration were performed as follows: (a) image stacking using the \textsc{reproject} package\footnote{\url{https://reproject.readthedocs.io/en/stable/index.html}}; (b) PSF photometry with \textsc{AutoPhOT}; and (c) flux calibration in the $BV$ and $gri$ bands using isolated AAVSO standard stars.

\subsubsection{Gemini South}
A series of imaging observations was obtained in $griz$ bands with the GMOS on Gemini South telescope \citep{Hook2004} at $\sim$31, 49, 79, and 88 days after the explosion under GS-2025A-DD-105
(PI. Wang). The data were pre-reduced using the DRAGONS package \citep{Labrie2023}\footnote{\href{https://www.gemini.edu/observing/phase-iii/reducing-data/dragons-data-reduction-software}{https://www.gemini.edu/observing/phase-iii/reducing-data/dragons-data-reduction-software}}, including bias subtraction, flat-field correction, bad-pixel masking, and mosaic coaddition. We then performed PSF and differential photometry using AUTOPHOT \citep{Brennan2022}. Most of the Gemini observations were calibrated against the skymapper catalog. However, in some $i$- and $z$-band images, some bright stars were saturated, making them unsuitable for photometric calibration with the relatively shallow skymapper catalog. In these cases, we adopted a catalog containing fainter stars and selected unsaturated faint stars as photometric references. We further used the Gaia--SDSS transformation relations from \citet{2018A&A...616A...4E} to transform the photometry into the SDSS $i$- 
and $z$-band system. The transformed catalog was used to determine the photometric zeropoints for the final two epochs of the $i$-band observations. 

An additional set of $gri$ images were acquired between July 7, 2026 and July 9, 2026 under GS-2026A-FT-218 (PI: Cikota) corresponding to 16 months post explosion. No source can be detected at the location of EP250304a/SN~2025fhm at a $3\sigma$ limiting magnitude of 24.7, 26.0, 25.9 for $gri$, respectively. 

\subsection{Optical spectroscopic observations}
\begin{figure}[htbp]
\centering
\includegraphics[width=\linewidth]{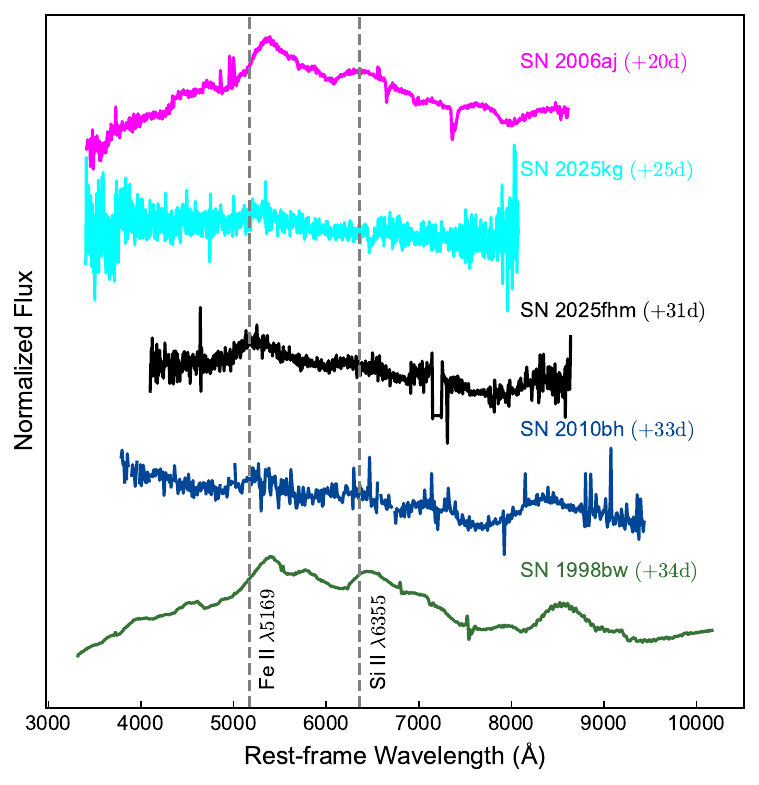}
\caption{The Gemini-GMOS optical spectrum of SN 2025fhm obtained at $\sim30$ days after the EP detection, shown alongside the comparison spectra of other SNe Ic-BL at similar epochs, including SN 2006aj \citep{Pian2006}, SN 2025kg \citep{Li_25_25kg}, SN 2010bh \citep{Bufano_12_10bh}, and SN 1998bw \citep{Patat_01_98bw}. The two broad features near $\sim 5000\AA$ and $6000\AA$ can be identified as Fe II $\lambda 5169$ and Si II $\lambda 6355$, respectively. Note the presence of a gap near 7000$\AA$ is due to the GMOS-S CCD failure. }
\label{fig:spectrum}
\end{figure}

We obtained a spectrum under GS-2025A-DD-105
(PI. Wang) for EP250304a/SN~2025fhm using the GMOS spectrograph on Gemini South on Apr 3rd, 2025 which is 31 days after its initial detection by EP (black line in figure \ref{fig:spectrum}). We triggered this spectrum with the R150/G5326 grating for a total of $3\times 1200$~s  exposure. The spectrum is then reduced using the IRAF Gemini/GMOS package for preprocessing and mosaicking the individual CCD chips into a single two-dimensional dispersed image. We then applied the standard IRAF spectroscopic reduction pipeline to perform spectral extraction, wavelength calibration using CuAr arc-lamp exposures, and flux calibration using the standard star LTT~7379.

The t$\sim$31 days spectrum (i.e., corresponding to $\sim$21 days post $BV$ peak and $\sim$16 days post $i$ peak) obtained for SN 2025fhm shows no detectable H or He emission lines, leading to a classification of Type Ic SN. In addition, the presence of broadened Fe II $\lambda5169$ and Si II $\lambda6355$ features is consistent with a broad-lined Type Ic (Ic-BL) classification. This proposed classification is further confirmed by template matching using SNID \citep{Blondin_07_SNID} and SNID-SAGE \citep{stoppa_2026_SNIDsage}. In particular, SN~2010bh at 14.8 days post-peak provides one of the best-matching templates at a redshift of $z \approx 0.22$. This template fitted redshift is consistent with $z = 0.2$ determined using host galaxy lines by \citet{GNC_spec_redshift}, and thus we adopt $z = 0.2$ in our following analysis.\newline
\indent In addition to SN~2010bh, we compare our spectrum with those of SN~2025kg \citep{Li_25_25kg}, SN~2006aj \citep{Pian2006}, and SN~1998bw \citep{Patat_01_98bw} at similar epochs. The spectrum of SN~2025fhm exhibits strong similarities to those of the SNe associated with prototypical low-luminosity GRBs.

\subsection{Radio observation}

Radio observations provide key information for probing the relativistic outflow of EP250304a and constraining its interaction with the circumburst medium. We conducted multi-frequency radio observations of EP250304a using the Australia Telescope Compact Array (ATCA) and the MeerKAT telescope. The observation log and flux measurements are summarized in Table 1.

Our monitoring campaign with the ATCA (at 5.5 and 9.0 GHz) started on Mar. 05 2025, approximately one day after the EP trigger. At these frequencies, we detected the radio emission with flux densities of $44 \pm 9$ $\mu$Jy and $81 \pm 8$ $\mu$Jy, respectively. Table 1 also includes two sets of measurements reported by other groups via Gamma-ray Coordination Network (GCN) Circulars. The first is the MeerKAT detection at 3.0 GHz on Mar. 07 2025 ($t \approx 3.13$ days), with a flux density of $\sim100$ $\mu$Jy, as reported by \cite{Carotenuto2025}. The second is the ATCA $3\sigma$ upper limit at 5.5 GHz on Mar. 21 2025 ($t \approx 17.6$ days), with a value of $<38$ $\mu$Jy, as reported by \cite{Yao2025}. All other measurements listed in the table are from this work.

\begin{table*}[htbp]
\centering
\caption{Radio observation log and flux measurements for EP250304a}
\label{tab:radio_EP250304a}
\begin{tabular}{l c c c c c c c}

\toprule
Observation Date (UT) & t (day) & Telescope & Frequency & Flux Density ($\mu$Jy) & Reference \\
\midrule
Mar 5, 2025 & 1 & ATCA & 5.5 GHz & $44 \pm 9$ & This work \\
Mar 5, 2025 & 1 & ATCA & 9.0 GHz & $81 \pm 8$ & This work \\
Mar 7, 2025  & 3.13 & MeerKAT & 3.0 GHz & $100 \pm 6.5$ & \cite{Carotenuto2025}  \\
Mar 11, 2025 & 7 & ATCA & 5.5 GHz & $<70$ & This work \\
Mar 11, 2025  & 7 & ATCA & 9.0 GHz & $<63$ &  This work \\
Mar 21, 2025 & 17.6 & ATCA & 5.5 GHz & $<38$ & \cite{Yao2025}  \\
\bottomrule
\end{tabular}
\end{table*}

\begin{figure}[htbp]
\centering
\includegraphics[scale=0.3]{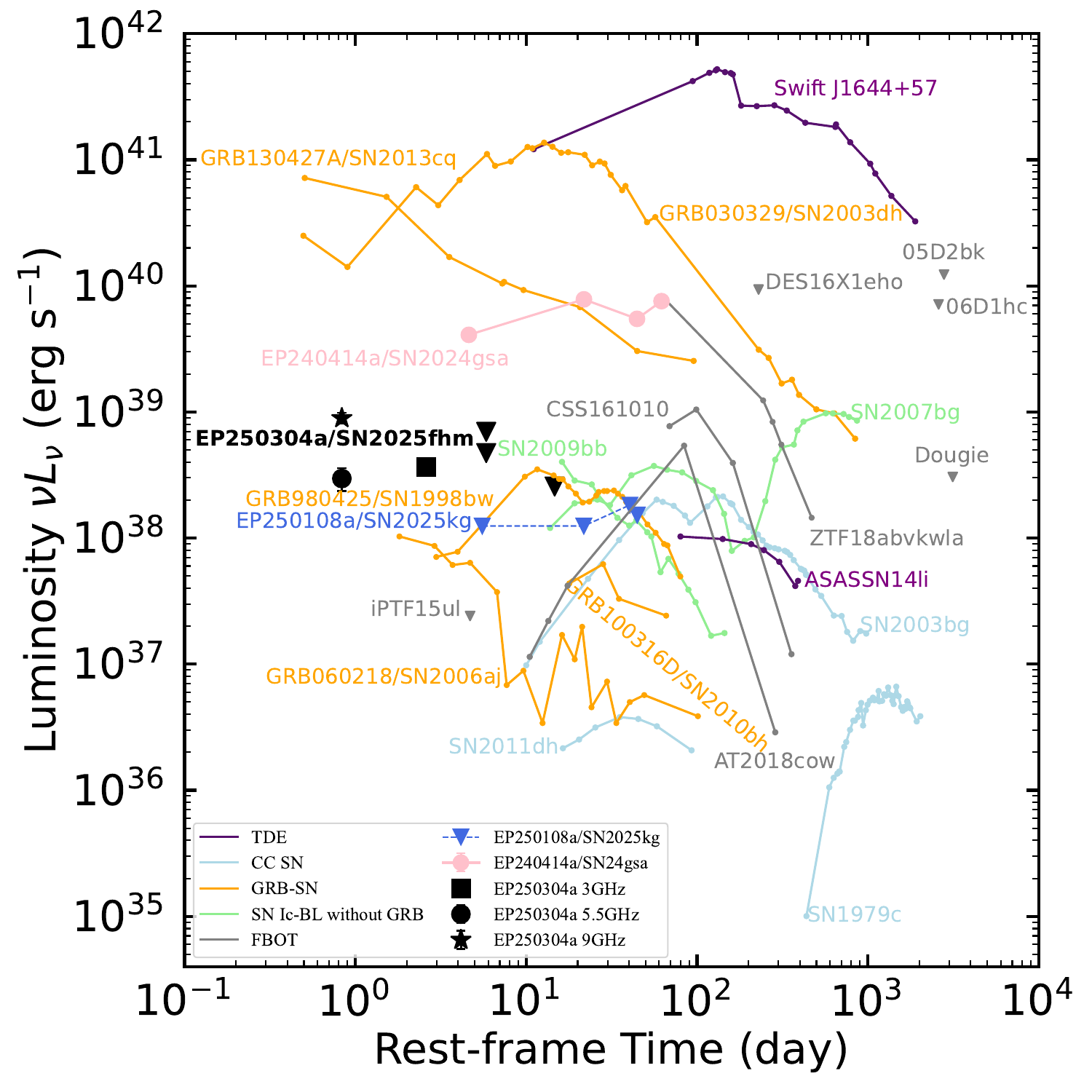}
\caption{Radio light curve of EP250304a at low frequencies (1–10 GHz), compared with those of various classes of energetic transients, including TDEs, CCSNe, relativistic SNe Ic-BL without GRB association, GRB-SNe, and FBOTs. }
\label{fig:radio}
\end{figure}

To place EP250304a/SN2025fhm in a broader context of high-energy transients, Figure~\ref{fig:radio} presents its radio luminosity upper limits at 3--10~GHz alongside the low-frequency (1--10~GHz) light curves of various transients from the literature, including TDEs, ordinary SNe, relativistic SNe Ic-BL, long GRBs, and FBOTs. The comparison shows that the radio upper limit of EP250304a/SN2025fhm lies significantly below those of typical long GRBs and most radio-detected TDEs, while being comparable to those of radio-quiet SNe Ic-BL and some radio-faint FBOTs. This might place the event within the category of bursts with mild relativistic outflows. Among EP-discovered FXT-SN events, EP250304a/SN2025fhm has a radio luminosity lying between the brighter EP240414a and the fainter EP250108a, indicating a continuous range of engine power or ambient density within this population. It is fainter than the SN-related GRBs like GRB~030329/SN~2003dh, but brighter than the SN-related low-luminosity GRBs like GRBs~060218/SN~2006aj and~980425/SN~1998bw.

This unique position highlights EP250304a/SN2025fhm as a key transitional object connecting classical GRBs, low-luminosity GRBs, and ordinary SNe Ic, likely powered by a jet of intermediate energy.

\section{Prompt and afterglow Emission}

The EP250304a X-ray light curve displays two broad pulses with comparable peak fluxes within the first $\sim$ 1 ks. The first pulse exhibits a clear hard-to-soft spectral index evolution, while the spectral index of the second pulse broadly follows the flux variations, indicative of an intensity-tracking pattern \citep{1983Natur.306..451G}. During the decay phase of the second pulse, the spectrum gradually hardens and then levels off at a near constant value of $\sim -2$. The corresponding temporal steep decay has an index of $\alpha_1=2.70_{-0.02}^{+0.02}$.

We found that the inclusion of a blackbody (BB) component improves the time-integrated spectral fitting to the first FXT epoch (see also \citet{2026arXiv260606213C}). The BB component exhibits spectral evolution during this epoch, which may be qualitatively consistent with thermal cooling emission from a jet-driven cocoon. The BB component is not detected at later epochs, suggesting that the subsequent X-ray emission is dominated by non-thermal emission from the jet and/or its afterglow.

Combining both the temporal and spectral slopes, we further tested whether the steep decay following the second pulse is consistent with the curvature effect \citep{2000ApJ...541L..51K, 2004ApJ...614..284D}, which arises when the prompt emission ceases abruptly and delayed high-latitude photons dominate the subsequent signal. In this scenario, the temporal and spectral indices satisfy the relation $\alpha=2+\beta$ \citep{2006ApJ...642..354Z, 2006ApJ...646..351L, 2009ApJ...690L..10Z}, where the indices are defined in the convention $F_\nu \propto t^{-\alpha} \nu^{-\beta}$ and converted from the spectral fitting index $\alpha_{\rm X}$ in Table \ref{tab:x_rays}. As illustrated in the inset of Figure \ref{fig:x_rays}, the measured temporal decay before the spectral hardening phase is shallower than the predicted slope, while after $\sim$1 ks it becomes broadly consistent with the model expectation. This behavior suggests that the steep decay is very likely the fading tail of the prompt emission, produced after the central engine shuts off and photons emitted from progressively higher latitudes relative to the line of sight arrive at later times. Subsequently, the light curve breaks to a shallow decay of $\alpha_2=0.41_{-0.16}^{+0.19}$ after $10^{4}$ s. Such a segment is consistent with the canonical X-ray afterglow evolution and could be interpreted within the standard external forward-shock framework, provided that sustained energy from a long-lived central engine or an impulsively ejected fireball with a stratified Lorentz factor continues to be injected into the blastwave.

Radio observations reveal a rapidly evolving emission region. On Day~1 ($t_{\text{rest}} \approx 7.2\times10^4$~s), ATCA detects the source at both 5.5~GHz and 9.0~GHz with an inverted spectrum ($F_\nu(9.0\ \text{GHz}) > F_\nu(5.5\ \text{GHz})$), placing the synchrotron self-absorption frequency $\nu_a$ above 9.0~GHz. Such a high $\nu_a$ signals a compact, relativistic shock with a brightness temperature $T_b \gg 10^{12}$~K, a defining hallmark of relativistic expansion. By Day~3.13 ($\approx 2.27\times10^5$~s), MeerKAT detects a strong signal at 3.0~GHz, with a flux density far exceeding the value extrapolated from the higher-frequency spectral shape on Day~1. This demonstrates that $\nu_a$ has dropped substantially from $>9.0$~GHz to near or below 3.0~GHz, roughly following a power-law decline $\nu_a \propto t^{-\delta}$ --- a behaviour naturally expected from adiabatic expansion of the shock. On Day~7 ($\approx 5.04\times10^5$~s), ATCA yields only non-detections at both 5.5~GHz and 9.0~GHz, providing stringent upper limits that imply the radio luminosity has decayed by more than an order of magnitude within one week. 

This evolutionary sequence imposes multiple constraints on the burst physics. The early high $\nu_a$ provides key evidence for a relativistic outflow. The rapid decline of $\nu_a$ traces the adiabatic cooling and expansion of the shock front, offering a handle on the initial kinetic energy of the jet and the density of the surrounding medium. 
The late non-detection, combined with the downward shift of $\nu_a$, points to a fast fading mechanism, possibly the shock entering a deceleration phase, the observing frequency falling into the optically thick regime, or the onset of a jet break. The radio evolution of EP250304a closely resembles that of classical low-luminosity GRB–SN systems (e.g., GRB 980425/SN 1998bw).

We perform fits to the data with conventional GRB afterglow models, where the emission is generally assumed to arise from a top-hat jet viewed on axis.
We fit the data using the open-source Python package \textsc{afterglowpy} \citep{Ryan2020}, now integrated into \textsc{Redback} \citep{Sarin2024}, which also provides a general class for refreshed afterglow models. Given the large number of model parameters and the limited X-ray and radio data available, we fix several parameters during the fitting: the ISM density $n_{\rm ISM} = 0.01\ {\rm cm}^{-3}$, the electron and magnetic energy fractions $\epsilon_e = 0.1$ and $\epsilon_B = 0.01$, the electron power-law index $p = 2.3$, and the acceleration fraction $\xi_N = 1$. The density-profile index $k$ is also fixed as $0$, corresponding to a constant-density medium \citep{Eldridge2007}. 
From our analysis, we infer a jet energy of $5.35_{-0.35}^{+0.43} \times 10^{50}\ {\rm erg}$, a jet opening angle of $0.27_{-0.09}^{+0.15}  \ {\rm rad}$, and an initial Lorentz factor of approximately $376.44_{-281.13}^{+417.02}$.

\begin{figure}[htbp]
\centering
\includegraphics[width=\linewidth]
{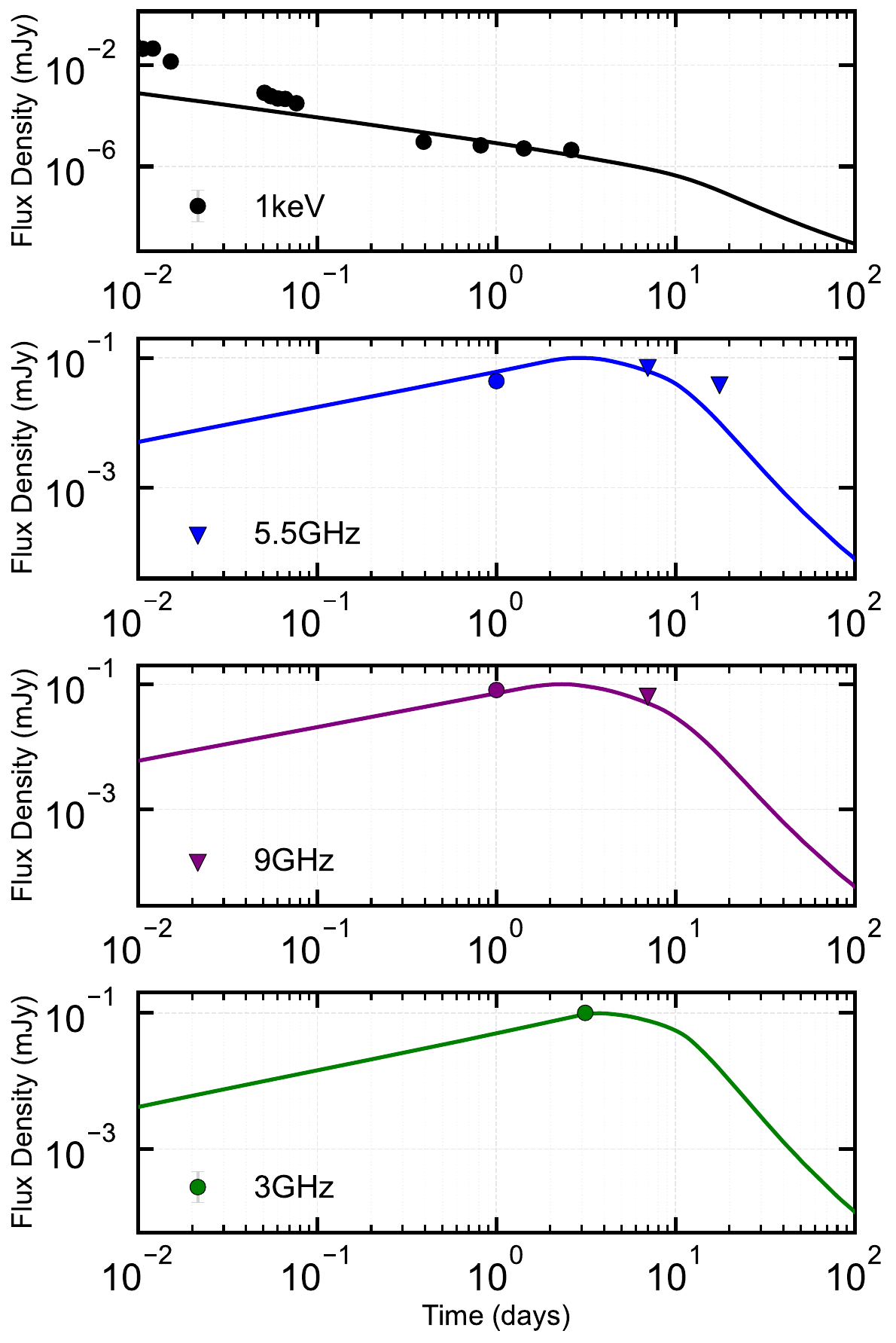}
\caption{The X-ray (1 keV) and radio light curves of EP250304a/SN 2025fhm, fitted with a top-hat jet afterglow model. }
\label{fig:X_radio_LC}
\end{figure}

\section{UV and Optical Emission}
\subsection{Light curve comparison}
\label{sec:lc_compare}

\begin{figure}[htbp]
\centering
\includegraphics[width=\linewidth]{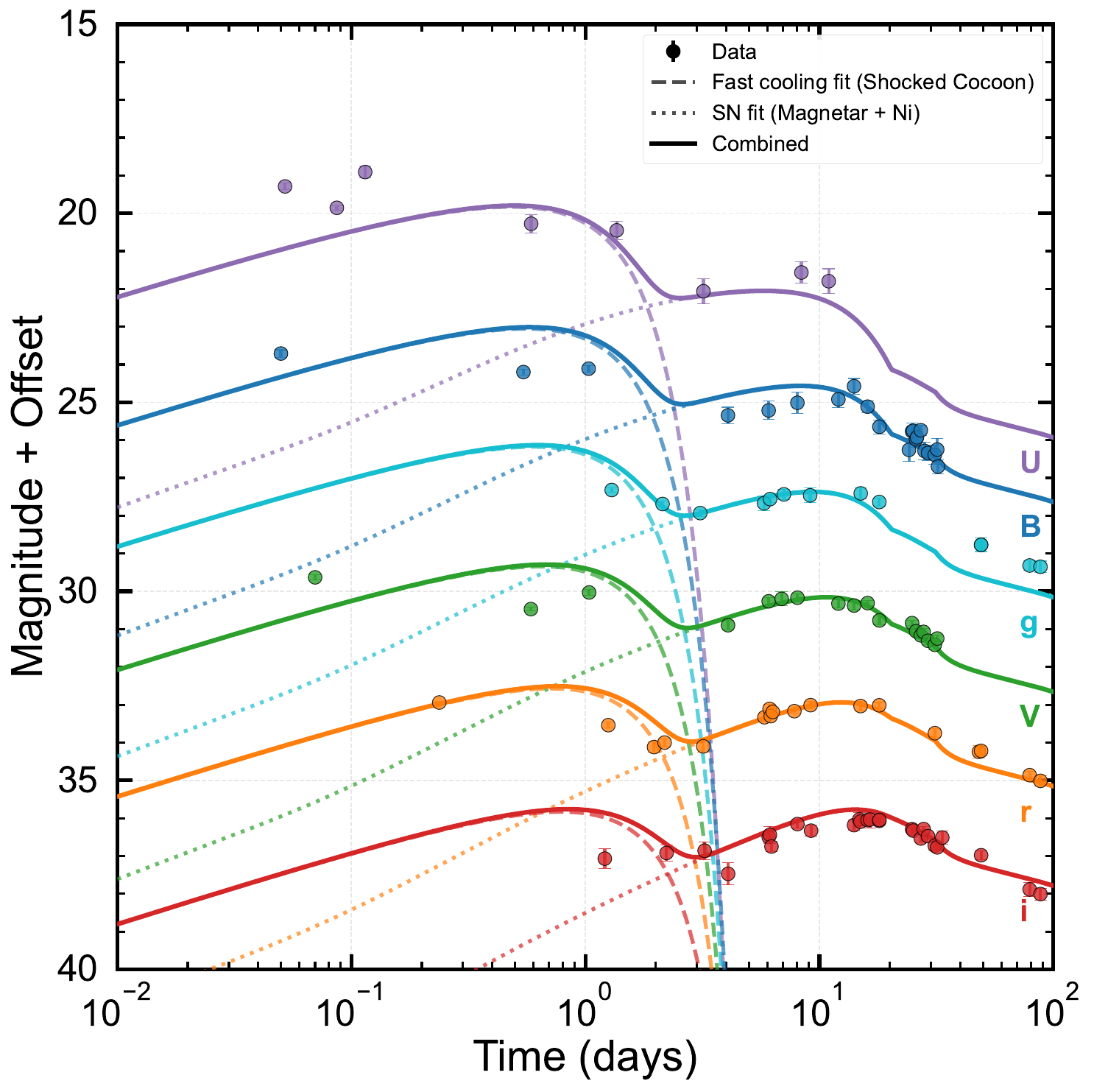}
\caption{Extinction-corrected multi-band light curves of optical counterpart for EP250304a (circles with error bars). EP250304a shows two distinctive phase of evolution: (1) A fast decline within first three days since discovery (fast-cooling phase) and (2) a bump that spans for 3--88 days since discovery (SN phase). The fast-cooling and SN phase are fitted with a shocked cocoon model (dashed lines) and combination of magnetar spin-down and $^\mathrm{56}$Ni radioactive decay (dotted lines), respectively. The resulting combined best-fit model is shown by the solid lines.}
\label{fig:multiband_magnetar}
\end{figure}

Figure \ref{fig:multiband_magnetar} shows the multi-band light evolution for the optical counterpart for EP250304a. The optical emission exhibits two distinct phases characterized by a) an initial rapid decline during the first $\sim$ 2--3 days post discovery, which we identify as the fast-cooling phase; b) and a subsequent bump spanning approximately 3--88 days post discovery, which we identify as the SN. 

We utilize the Automated Loss Regression pipeline \citep{Rodriguez_19_ALR} to perform non-parametric regression to our SN light curves for peak estimation. In the $BV$ bands, SN~2025fhm reaches maximum light at MJD = 60748, corresponding to $\sim$10 days after discovery, while the $i$-band peak occurs at MJD = 60756.34, or 18.27 days post-discovery. The inferred peak absolute magnitudes are $-18.1$ mag, $-18.9$ mag, and $-18.9$ mag in $BVi$ bands, respectively. Peak properties in the $g$ and $r$ bands are not estimated due to the lack of observations near the maximum light.

\begin{figure*}
\centering\includegraphics[width=\linewidth]{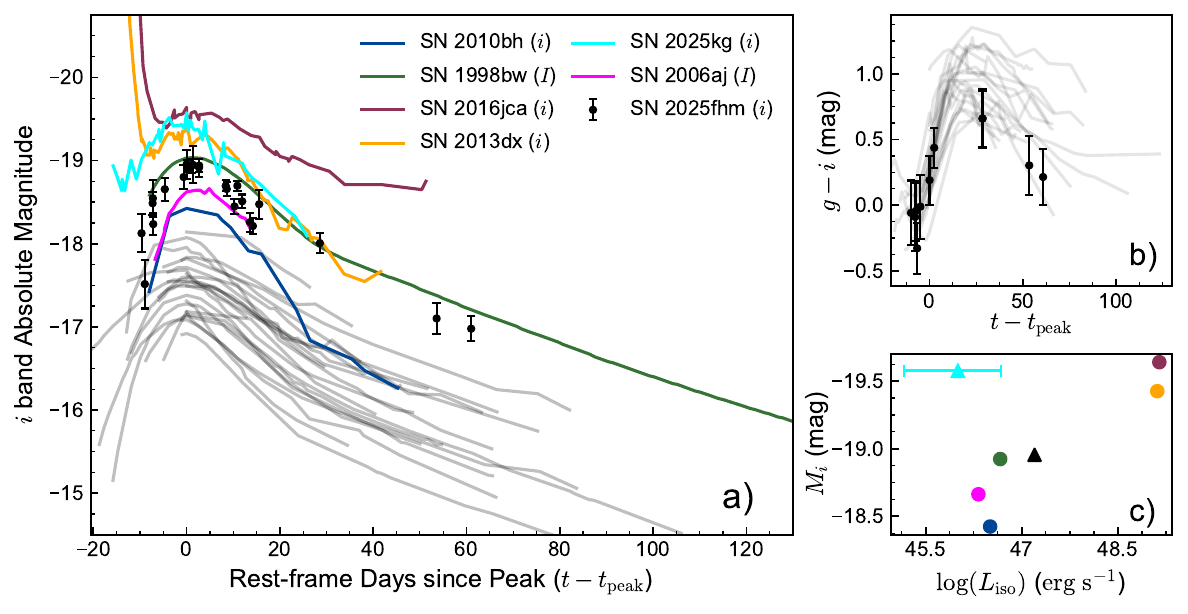}
    \caption{\textit{a):} The $i$-band light curve comparison between SN~2025fhm and SNe associated with high energy prompt emission including SN~2010bh \citep{Olivares_12_2010bh,Cano_11_10bh}, SN~1998bw \citep{Clocchiatti_11_98bw}, SN~2016jca \citep{Cano_17_16jca}, SN~2013dx \citep{Elia_15_13dx}, SN~2025kg \citep{Li_25_25kg}, and SN~2006aj \citep{Ferrero_06_06aj} and sample of SESNe from \citet[][grey lines]{Stritzinger_18_CSPSESNe}. Because the original light curves for SN~2006aj and SN~1998bw were observed in the $I$ band, we applied a transformation of $i=I+0.4$ to correct for the difference in zero points \citep{Park_17_ksp}. \textit{b):} Comparison of the extinction corrected $g-i$ colour between the 
    SN~2025fhm (filled circles with error bar) with the sample of SESNe from \citet[][grey lines]{Stritzinger_18_CSPSESNe}. \textit{c):} Comparison between the peak $i$ band magnitude of the SNe ($M_i$) with mean luminosity of the burst, defined to be $L_\mathrm{\gamma, iso}=E_\mathrm{\gamma, iso}(1+z)/T_{90}$ for the prompt emission, with $E_\mathrm{\gamma, iso}$ and $T_{90}$ denotes the isotropic equivalent emitted energy and the observed duration, respectively. Triangles indicate that source is first discovered by EP, while circles indicates the source is discovered by other facilities. The $L_\mathrm{\gamma, iso}$
    for SN2025kg is computed from the $T_{90}, E_\mathrm{\gamma, iso}$, and $z$ reported in \citet{Li_25_25kg}, while the rest are taken from \citet{Rudolph_22_Liso}. Note that we have taken $E_\mathrm{iso}$ to be $E_\mathrm{0.5-4\mathrm{KeV}}$ for both SN~2025kg and SN~2025fhm, given their ultra soft spectrum.}
    \label{fig:icomp}
\end{figure*}

Figure \ref{fig:icomp}a compares the $i$-band evolution of the optical counterpart with both the SESNe population \citep{Stritzinger_18_CSPSESNe} and a set of well-studied SN associated with high-energy prompt emission and good late-time coverage, including SN~2010bh \citep[][GRB 100316D]{Olivares_12_2010bh,Cano_11_10bh}, SN~1998bw \citep[][GRB 980425]{Clocchiatti_11_98bw}, SN~2016jca \citep[][GRB 161219B]{Cano_17_16jca}, SN~2013dx \citep[][GRB 130702A]{Elia_15_13dx}, SN~2025kg \citep[][EP250108A]{Li_25_25kg}, and SN~2006aj \citep[][GRB 060218]{Ferrero_06_06aj}.

The $i$-band light curve shows closer resemblance to those of GRB-SNe rather than the bulk of ordinary SESNe, despite the absence of a detected GRB counterpart. During the first $\sim$20 days after the peak, its evolution is most similar to that of SN~2006aj and SN~1998bw. In terms of luminosity, SN~2025fhm is brighter than SN~2010bh but remains less luminous than SN~2025kg, SN~2016jca, and SN~2013dx. At later epochs ($\gtrsim30$ days), SN~2025fhm continues to follow the decline observed in prototypical low-luminosity GRB-SNe such as SN~1998bw and SN~2013dx. While it remains fainter than SN~2016jca throughout its evolution \citep{Cano_17_16jca}. SN~2025fhm is systematically more luminous than the majority of the SESNe sample from \citet{Stritzinger_18_CSPSESNe}, as well as SN~2010bh during the nebular phase. Overall, its luminosity and temporal evolution place it much closer to the GRB-SN population than to the general SESN population.

In Figure~\ref{fig:icomp}b, we compare the extinction-corrected $g-i$ colour evolution of SN~2025fhm with those of the SESNe sample from \citet{Stritzinger_18_CSPSESNe}. At early epochs, SN~2025fhm follows closely the colour evolution of the SESNe population. However, at late times ($\gtrsim30$ days), it becomes significantly bluer than the bulk of the sample, as evident from the final three data points.

The limited sample shown in Figure~\ref{fig:icomp}c) tentatively suggests that more luminous prompt bursts may be associated with more luminous SNe on average. A similar trend has been reported by \citet{Rastinejad_25_25kg}, who find that higher X-ray peak luminosities tend to correspond to more luminous SNe. In this parameter space, SN~2025fhm occupies a region consistent with other low-luminosity GRBs, with similar X-ray and SNe luminosities. This may indicate a connection between prompt emission energetics and SN luminosity, although a significantly larger sample is required to test the robustness of this relationship.

\subsection{Modelling}
\subsubsection{Fast Cooling Phase}

The optical evolution of the counterpart to EP250304a closely resembles that observed in other LL-GRBs. Motivated by this similarity, we consider a scenario in which the initial fast-cooling phase is powered by emission from a shocked cocoon, as has been proposed for similar events (e.g., EP250108a \citealt{Eyles-Ferris_25_25kg}).

We model the early-phase data of the optical counterpart using the shocked cocoon model of \citealt{Prio2018}. In this framework, the early emission is attributed to cooling of material heated by a shock as driven by a relativistic outflow into the surrounding ejecta. Due to the limited observations during this phase, we fixed several parameters to typical and physically reasonable values. Specifically, we take the energy slope $\eta = 3.0$, consistent with the value inferred for shock-cooling events in \citealt{Prio2018}. The shocked mass fraction is set to $0.1$ and the cosine of the cocoon half-opening angle is $\cos\theta = 0.6$, corresponding to $\theta \approx 53^\circ$,and an opacity of $\kappa = 0.1\ {\rm cm^2\ g^{-1}}$. All these values are chosen to match the value used in our magnetar model for the late-time fitting, thereby maintaining a consistent physical treatment across both phases of the light curve.

As shown in Figure \ref{fig:multiband_magnetar}, the best shocked cocoon-fitting model (dashed lines), obtained through MCMC optimization, provides a decent match to the observed multicolor light curves. From this fitting, we derive the mass of the cocoon as $M_c = 0.14 \pm 0.02\ M_\odot$, the velocity of the cocoon as $v_c = 0.78 \pm 0.01\ c$, and the breakout time of the shock as $t_{\rm shock} = 20.7^{+2.1}_{-2.0} \rm ~s$, respectively. These values are physically reasonable for a supernova-associated outflow: the mass is a small fraction of the total ejecta, the velocity is mildly relativistic, and the shock breakout time of approximately 20 seconds is consistent with the short-lived cocoon emission phase.

\subsubsection{SN evolution}
We then explore the physical mechanisms that may power the SN evolution. SESNe are traditionally assumed to be powered solely by the radioactive decay of $^{56}\mathrm{Ni}$ and $^{56}\mathrm{Co}$, as described by Arnett’s model \citep{Arnett_82_Arnettlaw}. In this framework, the light-curve evolution is governed by radiative diffusion through the expanding ejecta, setting a characteristic timescale that depends on the ejecta mass, opacity, and expansion velocity, while the peak luminosity approximately traces the instantaneous heating rate from $^{56}\mathrm{Ni}$ decay at maximum light. However, this simplified picture may not capture the full diversity of SESN energy sources. Recent studies \citep[e.g.,][]{Rodriguez_24_centralengine} have suggested that additional energy injection from a central engine can also contribute to the observed luminosity evolution in at least some events.

\begin{figure*}
    \centering
    \includegraphics[width=\linewidth]{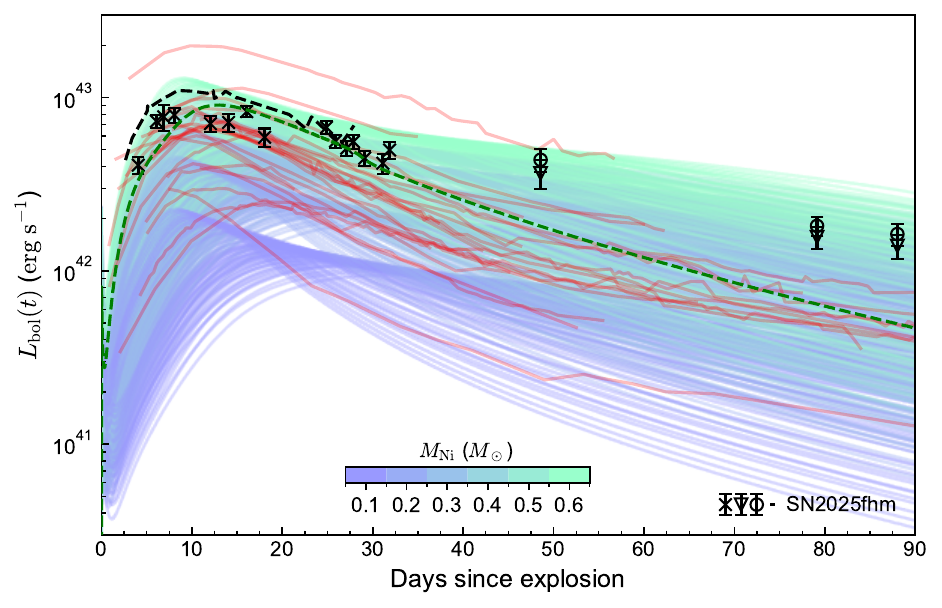}
    \caption{Comparison of the bolometric light curve of SN2025fhm with a sample of SESNe from \citet[][red curves]{Rodriguez_24_centralengine} and simulated light curves generated with the SNEC for a range of progenitor profiles and $^{56}$Ni masses (blue–green curves). The dashed green curve indicates the best fitted simulated model, with an explosion energy of 30B and a Ni mass of 0.5 $M_\odot$ mixed into the inner 90\% of the ejecta. For SN2025fhm, black dashed line, crosses, triangles, and circles denote bolometric luminosities derived using Superbol (under blackbody assumption), $Vi$, $gi$, and $ri$ corrections, respectively. For the simulated models, curve color encodes the $^{56}$Ni mass, as indicated by the colorbar.}
    \label{fig:SN_model}
\end{figure*}

To enable direct comparison with these models, we reconstruct the bolometric light curve of SN 2025fhm using \textit{Superbol} under the assumption of blackbody emission (see black dashed curve in Figure~\ref{fig:SN_model}). However, as the photosphere recedes into the inner ejecta, the emission is expected to deviate from local thermodynamic equilibrium, and the emitted spectrum is no longer well described by a blackbody. Therefore, we additionally reconstruct the bolometric luminosity using bolometric corrections from \citet{Rodriguez_23_bolometriccorrection} applied to our multi-band photometry. We adopt the $Vi$ bolometric corrections for epochs within 30 days of the discovery, as the dense KMTNet observations provide well-sampled coverage during this phase. At later times, owing to the lack of $V$-band data, we instead apply both $gi$ and $ri$ bolometric corrections to trace the late-time evolution. The bolometric luminosities derived from bolometric corrections is largely consistent with that estimated from \textit{Superbol}. 

We first investigate whether the light curve of SN~2025fhm can be explained solely by radioactive $^{56}$Ni decay by performing a grid of simulations using the 1D Lagrangian radiative hydrodynamics code SNEC \citep{SNEC_15,SNEC_16_early}. In all simulations, we fix the remnant neutron star mass to be 1.4$M_\odot$ and inject the explosion energy into the innermost 0.02$M_\odot$ of the progenitor over a duration of 1s, following a thermal bomb prescription. We adopt nine progenitor models from \citet{Song_23_GRBprogen}, spanning the progenitor masses of 1.7–7.2$M_\odot$. The explosion energies and $^{56}$Ni masses are varied over the ranges 3--10B and 0.1--0.6$M_\odot$, in increments of 1B and 0.1$M_\odot$, respectively, guided by previous modeling of \textit{ll}GRB-associated supernovae such as SN~2006aj \citep{Mazzali_06_2006aj}, SN~2023pel \citep{RomanAguilar_25_23pel}, and SN~2025kg \citep{Roman_Aguilar_25_25kg,Li_25_25kg}. Previous studies \citep{Taddia_19_IcBL,Fremling_26_supersnec} highlight the importance of $^{56}$Ni mixing in hydrodynamical modelling of SESNe. We therefore assume extensive mixing, distributing $^{56}$Ni throughout the inner 90\% of the ejecta, consistent with \citet{Taddia_19_IcBL}.

Figure~\ref{fig:SN_model} shows the comparison between the SNEC-predicted luminosities and the reconstructed bolometric light curve of SN~2025fhm. The best-fitting model in the explored parameter grid, corresponding to an explosion energy of $3\times10^{51}$~erg and a $^{56}$Ni mass of $0.5 M_\odot$ (green dashed line), provides a reasonable match to the observed luminosity evolution during the first $\sim$40 days since shock breakout. However, at epochs later than $\sim$50 days, the observed evolution became significantly brighter, consistent with the late-time colour behaviour noted in Section~\ref{sec:lc_compare}, where SN~2025fhm becomes significantly bluer than the comparison sample from \citet{Stritzinger_18_CSPSESNe}.  Relative to the sample of \citet[][red lines]{Rodriguez_24_centralengine}, SN~2025fhm remains more luminous than all comparison events at later evolution for ones showing similar early evolution, including SN~1998bw. This suggests that radioactive $^{56}$Ni decay alone is insufficient to fully account for the late-time luminosity evolution of SN~2025fhm.

Given that a pure Ni powered model fails to explain the late time evolution, we then explore a magnetar-powered model in which the central engine responsible for the prompt emission also injects energy into the SN ejecta. We perform multi-band light curve fitting using the semi-analytical magnetar model of \citet{Omand2024}, which includes energy injection from a rapidly spinning, highly magnetized neutron star that can supplement or dominate over radioactive decay at late times.

The model is fit to our multi-band data with the following free parameters: the initial spin-down luminosity $L_{0}$, the spin-down timescale $t_{\rm SD}$, the braking index $n$, the nickel mass fraction $f_{\rm Ni}$, the ejecta mass $M_{\rm ej}$, and the explosion energy $E_{\rm SN}$. In contrast to earlier works, \citet{Omand2024} parameterize the spin-down evolution in terms of $L_{0}$ and $t_{\rm SD}$ while allowing the braking index $n$ to vary. The canonical case $n=3$ corresponds to vacuum dipole spin-down. 

The best-fitting magnetar model is shown as the dashed curve in Figure~\ref{fig:multiband_magnetar}, while the corresponding combined cocoon+SN model is shown as the solid curve. The best fitted ejecta mass, $M_{\rm ej}=1.64_{-0.02}^{+0.03} M_\odot$, and explosion energy, $E_{\rm SN}=9.89_{-0.17}^{+0.08} \times 10^{51} \mathrm{erg~s^{-1}}$, lie within the typical range obtained for SNe Ic-BL. The nickel mass fraction is $f_{\rm Ni}=0.19\pm0.01$, indicating efficient $^{56}$Ni synthesis. The magnetar parameters are characterized by an initial spin-down luminosity of $L_{0}=9.85_{-0.23}^{+0.11} \times 10^{45} \mathrm{erg~s^{-1}}$ and a spin-down timescale of $t_{\rm SD}=1.4_{-0.04}^{+0.05} \times 10^{4} \rm~ s$. We then apply the scaling relations from Equations (21) and (22) of \citet{Omand_24_magnetarmodel} to infer an initial spin period as $\sim 12.60 \rm ~ms$ and a magnetic field strength as $\sim 3.52 \times 10^{15} \rm ~G$ for the magnetar.

\section{Discussion and Conclusion} 

This work presents a comprehensive multi-wavelength analysis and modelling of EP250304a, a fast X-ray transient discovered by the EP, and its optical counterpart, the broad-lined Type Ic supernova SN~2025hfm. By systematically integrating data from X-ray, optical, and radio bands, and performing combined fits with afterglow and supernova engine models, we construct a unified physical picture of this event and place it within the broader evolutionary framework of high-energy astrophysical transients.

Analysis of the late-time supernova light curve reveals the presence of energy injection, suggesting that the central engine powering the burst is likely a magnetar. The standard radioactive nickel decay (Arnett) model cannot account for the anomalously shallow decay slope ($\alpha_{\mathrm{opt}} \approx 0.2$) observed around $\sim40$ days post-peak, whereas a magnetar-powered model naturally resolves this issue, consistently pointing to a newborn, highly magnetised, rapidly rotating magnetar. Furthermore, the early afterglow evolution deviates from the standard afterglow model, suggesting that additional physical processes are at work. This is consistent with ongoing energy injection from a long-lived central engine such as a magnetar. 

As a central engine candidate, the magnetar has been extensively studied in the contexts of long GRBs, GRB-associated SNe, and SLSNe over the past two decades \citep[e.g.,][]{Zhang2001, Woosley2010, Metzger2011,Nicholl2017,Margalit2018}. For superluminous supernovae (SLSNe), the magnetar engine model has received strong observational support from recent events such as SN\,2024afav, whose quasi-periodic light-curve modulations point to a precessing magnetar with a fallback disk \citep{Farah2026}, and SN\,2017egm, whose GeV emission matches theoretical expectations \citep{Li2026}. On the statistical side, a comprehensive study modeling the light curves of 15 GRB-associated or relativistic-kicked supernovae derived the median engine parameters: an ejecta mass of $\sim 5.2\,M_\odot$, an initial spin period of $\sim 20.5$~ms, and a magnetic field strength of $\sim 2.01\times10^{15}$~G \citep{Kumar2024}. These derived parameters firmly place the sample within the magnetar regime and demonstrate that variations in the initial period and magnetic field strength can naturally produce diverse explosive transients, ranging from GRB-SNe to SLSNe and even FBOTs.

Within the statistical sample of FXTs and SNe~Ic-BL, EP250304a, together with EP240414a and EP250108a, forms a distinct subclass. Its properties lie between high-luminosity classical GRB-SNe and low-luminosity GRBs, while also resembling those relativistic SNe without GRB associations, such as SN 2012ap and SN 2009bb \citep[e.g.,][]{Pignata2011,Margutti2014,Chakraborti2015,Modjaz2016}. Notably, SN 2012ap and SN 2009bb show evidence of relativistic outflows, yet neither GRBs nor FXTs have been detected in association with them. It remains possible that these events do produce X-ray emission, but previous instruments lacked the sensitivity to detect them. The EP with its unique wide-field monitoring capabilities, is ideally suited to uncover such hidden X-ray counterparts, positioning EP250304a as a key transitional object bridging classical GRBs, low luminosity GRBs, relativistic SNe lacking high-energy emission, and fully ``silent'' engine-driven supernovae. Future EP detections are expected to greatly expand this sample, revealing a continuous physical sequence in engine power and outflow collimation.

\begin{table*}
\centering
\scriptsize
\caption{Spectral fitting results and corresponding fitting statistics for EP-WXT and FXT. All errors represent the 1$\sigma$ uncertainties.}
\label{tab:x_rays}
\begin{tabular}{cccccc}
\hline
\hline
\multirow{3}{*}{t1 (s)} & \multirow{3}{*}{t2 (s)} & \multicolumn{3}{c}{PL model} & \multirow{2}{*}{$F_\nu$ (1 keV)}\\
\cline{3-5}
& & $\alpha$ & log$A$ & stat/dof & \\
& & & ($\rm{photons~cm^{-2}~s^{-1}~keV^{-1}}$) & & (erg s$^{-1}$ cm$^{-2}$ keV$^{-1}$) \\
\hline
2.5 & 40.5 & ${-1.01}_{-0.57}^{+0.48}$ & ${-0.65}_{-0.16}^{+0.09}$ & 0.828/3 & ${3.59}_{-1.12}^{+0.82} \times 10^{-10}$ \\
40.5 & 61.0 & ${-1.69}_{-0.62}^{+0.56}$ & ${-0.36}_{-0.17}^{+0.09}$ & 2.679/3 & ${6.95}_{-2.24}^{+1.52} \times 10^{-10}$ \\
61.0 & 74.0 & ${-1.83}_{-0.70}^{+0.54}$ & ${-0.14}_{-0.16}^{+0.09}$ & 4.875/3 & ${1.15}_{-0.35}^{+0.25} \times 10^{-9}$ \\
74.0 & 85.5 & ${-2.41}_{-0.95}^{+0.61}$ & ${-0.13}_{-0.16}^{+0.08}$ & 3.101/2 & ${1.18}_{-0.37}^{+0.23} \times 10^{-9}$ \\
85.5 & 99.5 & ${-2.89}_{-0.98}^{+0.59}$ & ${-0.24}_{-0.18}^{+0.08}$ & 1.372/2 & ${9.22}_{-3.11}^{+1.77} \times 10^{-10}$ \\
99.5 & 119.0 & ${-2.79}_{-0.94}^{+0.61}$ & ${-0.37}_{-0.20}^{+0.08}$ & 6.037/3 & ${6.85}_{-2.49}^{+1.42} \times 10^{-10}$ \\
119.0 & 140.0 & ${-2.42}_{-1.14}^{+0.56}$ & ${-0.39}_{-0.22}^{+0.08}$ & 2.129/3 & ${6.59}_{-2.62}^{+1.35} \times 10^{-10}$ \\
140.0 & 167.5 & ${-2.15}_{-0.68}^{+0.70}$ & ${-0.48}_{-0.19}^{+0.08}$ & 1.065/3 & ${5.28}_{-1.86}^{+1.09} \times 10^{-10}$ \\
167.5 & 194.0 & ${-3.16}_{-0.84}^{+0.60}$ & ${-0.56}_{-0.19}^{+0.10}$ & 1.378/3 & ${4.42}_{-1.55}^{+1.09} \times 10^{-10}$ \\
194.0 & 217.5 & ${-2.91}_{-0.80}^{+0.59}$ & ${-0.48}_{-0.15}^{+0.09}$ & 3.079/5 & ${5.35}_{-1.57}^{+1.26} \times 10^{-10}$ \\
217.5 & 252.0 & ${-2.50}_{-0.50}^{+0.44}$ & ${-0.65}_{-0.10}^{+0.07}$ & 4.633/8 & ${3.61}_{-0.72}^{+0.62} \times 10^{-10}$ \\
252.0 & 285.5 & ${-2.07}_{-0.44}^{+0.31}$ & ${-0.60}_{-0.08}^{+0.05}$ & 12.180/13 & ${4.06}_{-0.70}^{+0.51} \times 10^{-10}$ \\
285.5 & 306.4 & ${-2.08}_{-0.40}^{+0.32}$ & ${-0.39}_{-0.08}^{+0.05}$ & 11.857/13 & ${6.48}_{-1.13}^{+0.76} \times 10^{-10}$ \\
\hline
296.0 & 313.5 & ${-2.20}_{-0.12}^{+0.11}$ & ${-0.33}_{-0.03}^{+0.02}$ & 9.908/17 & ${7.48}_{-0.54}^{+0.36} \times 10^{-10}$ \\
313.5 & 328.0 & ${-2.21}_{-0.14}^{+0.12}$ & ${-0.23}_{-0.03}^{+0.02}$ & 9.573/18 & ${9.37}_{-0.60}^{+0.45} \times 10^{-10}$ \\
328.0 & 343.0 & ${-2.18}_{-0.14}^{+0.12}$ & ${-0.26}_{-0.03}^{+0.03}$ & 15.609/18 & ${8.80}_{-0.56}^{+0.52} \times 10^{-10}$ \\
343.0 & 356.5 & ${-2.29}_{-0.12}^{+0.14}$ & ${-0.23}_{-0.03}^{+0.02}$ & 13.728/16 & ${9.52}_{-0.68}^{+0.47} \times 10^{-10}$ \\
356.5 & 368.5 & ${-2.25}_{-0.12}^{+0.11}$ & ${-0.18}_{-0.03}^{+0.02}$ & 31.233/18 & ${1.06}_{-0.07}^{+0.06} \times 10^{-9}$ \\
368.5 & 380.0 & ${-2.41}_{-0.13}^{+0.11}$ & ${-0.16}_{-0.03}^{+0.02}$ & 30.365/18 & ${1.11}_{-0.07}^{+0.06} \times 10^{-9}$ \\
380.0 & 391.5 & ${-2.44}_{-0.13}^{+0.11}$ & ${-0.14}_{-0.03}^{+0.02}$ & 12.334/17 & ${1.17}_{-0.08}^{+0.06} \times 10^{-9}$ \\
391.5 & 403.5 & ${-2.34}_{-0.11}^{+0.11}$ & ${-0.17}_{-0.03}^{+0.02}$ & 14.967/17 & ${1.08}_{-0.07}^{+0.05} \times 10^{-9}$ \\
403.5 & 416.0 & ${-2.40}_{-0.15}^{+0.14}$ & ${-0.17}_{-0.03}^{+0.02}$ & 22.250/19 & ${1.07}_{-0.07}^{+0.06} \times 10^{-9}$ \\
416.0 & 428.0 & ${-2.59}_{-0.11}^{+0.11}$ & ${-0.17}_{-0.03}^{+0.02}$ & 21.054/17 & ${1.09}_{-0.07}^{+0.06} \times 10^{-9}$ \\
428.0 & 439.5 & ${-2.48}_{-0.11}^{+0.12}$ & ${-0.17}_{-0.03}^{+0.03}$ & 26.509/16 & ${1.08}_{-0.07}^{+0.07} \times 10^{-9}$ \\
439.5 & 453.0 & ${-2.51}_{-0.12}^{+0.14}$ & ${-0.22}_{-0.03}^{+0.02}$ & 17.544/18 & ${9.57}_{-0.64}^{+0.50} \times 10^{-10}$ \\
453.0 & 466.5 & ${-2.70}_{-0.13}^{+0.13}$ & ${-0.21}_{-0.04}^{+0.02}$ & 8.159/16 & ${9.88}_{-0.75}^{+0.50} \times 10^{-10}$ \\
466.5 & 482.0 & ${-3.17}_{-0.12}^{+0.15}$ & ${-0.29}_{-0.03}^{+0.02}$ & 21.328/18 & ${8.12}_{-0.59}^{+0.42} \times 10^{-10}$ \\
482.0 & 501.0 & ${-2.98}_{-0.13}^{+0.14}$ & ${-0.38}_{-0.03}^{+0.03}$ & 13.708/17 & ${6.76}_{-0.44}^{+0.41} \times 10^{-10}$ \\
501.0 & 519.5 & ${-2.70}_{-0.14}^{+0.13}$ & ${-0.37}_{-0.03}^{+0.03}$ & 27.259/17 & ${6.80}_{-0.43}^{+0.41} \times 10^{-10}$ \\
519.5 & 541.5 & ${-3.13}_{-0.13}^{+0.13}$ & ${-0.45}_{-0.03}^{+0.02}$ & 14.495/17 & ${5.71}_{-0.40}^{+0.32} \times 10^{-10}$ \\
541.5 & 567.0 & ${-2.92}_{-0.14}^{+0.13}$ & ${-0.51}_{-0.03}^{+0.02}$ & 18.915/17 & ${4.99}_{-0.32}^{+0.26} \times 10^{-10}$ \\
567.0 & 595.5 & ${-3.07}_{-0.15}^{+0.18}$ & ${-0.58}_{-0.03}^{+0.02}$ & 30.149/16 & ${4.21}_{-0.30}^{+0.25} \times 10^{-10}$ \\
595.5 & 627.5 & ${-2.95}_{-0.18}^{+0.19}$ & ${-0.63}_{-0.03}^{+0.02}$ & 24.843/16 & ${3.76}_{-0.26}^{+0.18} \times 10^{-10}$ \\
627.5 & 662.5 & ${-2.97}_{-0.17}^{+0.18}$ & ${-0.67}_{-0.03}^{+0.02}$ & 24.799/17 & ${3.46}_{-0.27}^{+0.16} \times 10^{-10}$ \\
662.5 & 709.0 & ${-2.70}_{-0.20}^{+0.19}$ & ${-0.77}_{-0.03}^{+0.02}$ & 19.817/17 & ${2.71}_{-0.19}^{+0.14} \times 10^{-10}$ \\
709.0 & 772.0 & ${-2.73}_{-0.20}^{+0.21}$ & ${-0.92}_{-0.03}^{+0.02}$ & 28.425/16 & ${1.92}_{-0.14}^{+0.10} \times 10^{-10}$ \\
772.0 & 859.5 & ${-2.96}_{-0.20}^{+0.18}$ & ${-1.07}_{-0.03}^{+0.02}$ & 27.404/16 & ${1.35}_{-0.10}^{+0.08} \times 10^{-10}$ \\
859.5 & 974.0 & ${-2.49}_{-0.15}^{+0.16}$ & ${-1.18}_{-0.03}^{+0.02}$ & 26.732/17 & ${1.06}_{-0.08}^{+0.05} \times 10^{-10}$ \\
974.0 & 1150.5 & ${-2.13}_{-0.19}^{+0.17}$ & ${-1.40}_{-0.03}^{+0.02}$ & 27.353/16 & ${6.33}_{-0.48}^{+0.36} \times 10^{-11}$ \\
1150.5 & 1445.0 & ${-1.97}_{-0.15}^{+0.17}$ & ${-1.57}_{-0.03}^{+0.02}$ & 23.896/19 & ${4.33}_{-0.33}^{+0.21} \times 10^{-11}$ \\
4171.0 & 4523.5 & ${-1.77}_{-0.16}^{+0.14}$ & ${-2.94}_{-0.04}^{+0.03}$ & 8.361/12 & ${1.86}_{-0.14}^{+0.13} \times 10^{-12}$ \\
4523.5 & 4946.5 & ${-1.85}_{-0.17}^{+0.15}$ & ${-3.02}_{-0.04}^{+0.03}$ & 16.946/12 & ${1.52}_{-0.12}^{+0.11} \times 10^{-12}$ \\
4946.5 & 5425.0 & ${-1.93}_{-0.19}^{+0.16}$ & ${-3.11}_{-0.04}^{+0.03}$ & 12.614/11 & ${1.25}_{-0.11}^{+0.08} \times 10^{-12}$ \\
5425.0 & 6041.0 & ${-1.98}_{-0.14}^{+0.14}$ & ${-3.18}_{-0.03}^{+0.03}$ & 15.623/13 & ${1.06}_{-0.08}^{+0.07} \times 10^{-12}$ \\
6041.0 & 7205.0 & ${-1.82}_{-0.12}^{+0.14}$ & ${-3.30}_{-0.03}^{+0.02}$ & 17.128/19 & ${8.09}_{-0.61}^{+0.43} \times 10^{-13}$ \\
32977.0 & 34847.0 & -2 & ${-4.96}_{-0.33}^{+0.12}$ & 8.194/8 & ${1.78}_{-0.95}^{+0.57} \times 10^{-14}$ \\
67543.0 & 73603.0 & -2 & ${-4.89}_{-0.11}^{+0.07}$ & 20.435/21 & ${2.06}_{-0.47}^{+0.36} \times 10^{-14}$ \\
119392.0 & 126454.0 & -2 & ${-5.05}_{-0.13}^{+0.08}$ & 32.279/20 & ${1.41}_{-0.36}^{+0.28} \times 10^{-14}$ \\
223087.0 & 229135.0 & -2 & ${-5.27}_{-0.36}^{+0.11}$ & 10.907/14 & ${8.66}_{-4.92}^{+2.34} \times 10^{-15}$ \\
309498.0 & 316217.0 & -2 & ${-5.27}_{-0.27}^{+0.11}$ & 14.596/18 & ${8.53}_{-3.90}^{+2.53} \times 10^{-15}$ \\
932000.0$^*$ & 933683.0$^*$ & - & - & - & ${1.49} \times 10^{-14}$ \\
1279500.0$^*$ & 1287072.0$^*$ & - & - & - & ${1.35} \times 10^{-14}$ \\
\hline
\hline
\end{tabular}
\begin{tablenotes}
\scriptsize
\item * upper limits.
\end{tablenotes}
\end{table*}

\begin{acknowledgments}
We thank the many helpful conversations with He Gao, Tong Liu, Tianrui Sun, Conor Omand, Anna Ho, Ehud Nakar, Maria Drout, Jianing Su and Chris Irwin.

This work is supported by the National Natural Science Foundation of China (NSFC grants 12288102, 12033003, 11633002,12573050), the Tencent Xplorer Prize, Scientific Research Foundation of Hainan Tropical Ocean University (NO. RHDRCZK202622), the China Postdoctoral Science Foundation (NO.2024T170455, GZC20231350, 2023M741998), the Chinese Academy of Sciences South America Center for Astronomy (CASSACA) Key Research Project E52H540301, and in part by the Chinese Academy of Sciences (CAS) through a grant to the CASSACA. WXL is supported by the NSFC (Grant Nos. 12120101003 and 12373010), the National Key R$\&$D Program of China (Grant Nos. 2022YFA1602902 and 2023YFA1607804), the Strategic Priority Research Program of the Chinese Academy of Sciences (Grant Nos. XDB0550100 and XDB0550000), and the NAOC Programs (Grant Nos. E5ZB7801). This research is also based on observations obtained at the international Gemini-S Observatory, a program of National Science Foundation’s (NSF) NOIRLab, which is managed by the Association of Universities for Research in Astronomy (AURA) under a cooperative agreement with the NSF on behalf of the Gemini Observatory partnership: the NSF (United States), National Research Council (NRC; Canada), Agencia Nacional de Investigaci\'{o}n y Desarrollo (Chile), Ministerio de Ciencia, Tecnolog\'{i}a e Innovaci\'{o}n (Argentina), Minist\'{e}rio da Ci\^{e}ncia, Tecnologia, Inova\c{c}\~{o}es e Comunica\c{c}\~{o}es (MCTI; Brazil), and KASI (Republic of Korea). Archive at NSF’s NOIRLab. This work makes use of observations from the Las Cumbres Observatory network. The LCO team is supported by NSF grant AST-2308113. We thank the Global Supernova Project collaboration. This research is based on observations made with the TRT under program ID $\rm TRTC12A\_003$, which is operated by the National Astronomical Research Institute of Thailand (Public Organization). 

This research has made use of the KMTNet system operated by the Korea Astronomy and Space Science Institute (KASI) and the data were obtained at three host sites of CTIO in Chile, SAAO in South Africa, and SSO in Australia. Data transfer from the host site to KASI was supported by the Korea Research Environment Open NETwork (KREONET). This research was supported by KASI under the R\&D program (Project No. 2026-1-831-02), supervised by the Korea AeroSpace Administration. D.-S.M. is supported by Discovery Grants from the Natural Sciences and Engineering Research Council of Canada (NSERC; Nos. RGPIN-2019-06524). D.-S.M. was supported in part by a Leading Edge Fund from the Canadian Foundation for Innovation (CFI; project No. 30951).  CDM is supported by an NSERC Discovery Grant. 
FEB acknowledges support from ANID-Chile BASAL
CATA FB210003 and FONDECYT Regular 1241005.

\end{acknowledgments}

\begin{contribution}
Cui-Ying Song and Nan Jiang contributed equally to this work, leading the data analysis, interpretation, and manuscript writing. Xiaofeng Wang and Wenxiong Li conceived and supervised the project, coordinated the collaboration. Yi-Han Iris Yin analyzed the EP X-ray data and wrote the corresponding text. Lingzhi Wang, and Shengyu Yan obtained and processed the Gemini observational data. Samaporn Tinyanont, and Ning-Chen Sun responsible for the proposal and acquisition of the TRT observational data, as well as the subsequent data processing. D. Andrew Howell supplied the LCO observational data. An Tao contribute to radio observational data. All authors reviewed and approved the final manuscript.

\end{contribution}

\facilities{EP, Swift (XRT and UVOT), Ferm, KMTNet, TRT, LCO, GMOS on Gemini South Telescope, the Global Supernova Network, ATCA, MeerKAT. }

\software{astropy \citep{Astropy2013,Astropy2018,Astropy2022}, afterglowpy \citep{Ryan2020}, Superbol \citep{Nicholl2018}, Redback \citep{Sarin2024}, Mesa \citep{Paxton2011,Paxton2013,Paxton2015,Paxton2018,Paxton2019,Jermyn2023}, SNEC \citep{Morozova2015}, DRAGONS package \citep{Labrie2023}, AUTOPHOT \citep{Brennan2022}.}

\appendix

\bibliography{ms}{}
\bibliographystyle{aasjournalv7}

\end{document}